\documentclass[aps,prx,showpacs,preprintnumbers,twocolumn,superscriptaddress]{revtex4-2}
\usepackage{amsmath,amssymb}
\usepackage{bm}
\usepackage{tipa}
\usepackage{upgreek}
\usepackage{comment}
\usepackage{mathrsfs}
\usepackage{graphicx}
\usepackage{braket}
\usepackage{leftindex}
\usepackage{enumitem}
\usepackage{mathbbol}
\usepackage{gensymb}
\usepackage[normalem]{ulem}
\usepackage{color}
\usepackage[colorlinks,bookmarks=true,citecolor=blue,linkcolor=red,urlcolor=blue]{hyperref}
\usepackage{hyperref}
\usepackage{dsfont}

\begin{document}

\title{Cross-platform protected qubits from entanglement}
\begin{abstract}
A crucial ingredient for scalable fault-tolerant quantum computing is the construction of logical qubits with low error rates and intrinsic noise protection. We propose a \emph{cross-platform} construction for such hardware-level noise-protection in which the qubits are protected from depolarizing (relaxation) \textit{and} dephasing errors induced by local noise. These logical qubits arise from the entanglement between two internal degrees of freedom, hence we term them \textit{entanglemons}. Our construction is based on the emergence of collective degrees of freedom from a generalized coherent state construction, similar in spirit to spin coherent states, of a set of such internally entangled units. These degrees of freedom, for a finite number of units, parametrize the quantized version of complex projective space $\mathbb{C}$P(3). The noise protection of the entanglemon qubit is then a consequence of a weakly coupled emergent degree of freedom arising due to the non-linear geometry of complex projective space. We present two simple models for entanglemons which are platform agnostic, provide varying levels of protection and in which the qubit basis states
are the two lowest energy states with a higher energy gap to other states. We end by commenting on how entanglemons could be realized in platforms ranging from superconducting circuits and trapped ion platforms to possibly also quantum Hall skyrmions in graphene and quantum dots in semiconductors.
The inherent noise protection in our models combined with the platform agnosticism highlights the potential of encoding information in additional weakly coupled emergent degrees of freedom arising in non-linear geometrical spaces and curved phase spaces, thereby proposing a different route to achieve scalable fault-tolerance.
\end{abstract}
\author{Nilotpal Chakraborty}
\email{nilotpal@pks.mpg.de}
\affiliation{Max-Planck-Institut f\"{u}r Physik komplexer Systeme, N\"{o}thnitzer Stra\ss e 38, Dresden 01187, Germany}

\author{Roderich Moessner}
\affiliation{Max-Planck-Institut f\"{u}r Physik komplexer Systeme, N\"{o}thnitzer Stra\ss e 38, Dresden 01187, Germany}

\author{Benoit Doucot}
\affiliation{LPTHE, UMR 7589, CNRS and Sorbonne Universit\'e, 75252 Paris Cedex 05, France}

\maketitle

\section{Introduction}
Building a robust universal quantum computer is a great interdisciplinary challenge \cite{nielsen2010quantum}. One of the key obstacles in this pursuit is dealing with the  sensitivity of qubits to even the slight amounts of noise. The ultimate goal of fault-tolerant quantum computation (FTQC) relies on the most basic computational elements being robust to errors induced by the unavoidable noise \cite{shor1996fault,preskill1998fault,Gottesmann}. Proposals to achieve this goal have been at the forefront of quantum computing research over the last two decades \cite{bravyi1998quantum,dennis2002topological,fowlersurf,kitaev2003fault}. The pursuit of FTQC takes two distinct routes - i) Identifying errors and performing quantum error correction (QEC) \cite{qecknill,gottesman1997stabilizer}, a field which has shown remarkable recent theoretical and experimental progress in various platforms from superconducting circuits to Rydberg atoms and ion traps \cite{bluvstein2024logical,xu2024constant,bravyi2024high}. ii) Building qubits which are inherently protected to some or all kinds of noise. Proposals such as Kitaev's topological quantum computing using non-abelian anyons \cite{kitaev2003fault}, the Gottesmann-Kitaev-Preskill (GKP) qubit \cite{gottesman2001encoding} or even the more recent transmon \cite{koch2007charge}, fluxonium \cite{manufluxonium} and $0-\pi$ qubits \cite{brooks0pi} in superconducting circuits represent notable advances in this category \cite{beytransm}. The final goal of  FTQC will likely require a combination of ideas from both categories.

In this work, we are motivated by proposals in the latter category. We propose a \emph{cross-platform} construction for qubits protected from depolarization \emph{and} dephasing errors induced by local noise.  Under the simplest uniform noise estimates, our qubit construction has no non-zero matrix element for any local noise operator (we comment on more realistic estimates at the end). Our construction invokes the idea of local entanglement between physical but internal degrees of freedom, hence we call the resulting qubits \emph{entanglemons}. These internal degrees of freedom could be a spin and valley degree of freedom in solid-state platforms such as graphene or silicon, electronic spin and nuclear spin in hyperfine ion traps or two plaquettes in a Josephson junction array. The qubit corresponding to the entanglement phase $\beta$, arises from a collection of such internally entangled units. 

Prior to discussing concrete realizations, we present two simple models for such entanglemons, to  provide an understanding of their working principle. The first, model $U(1)^{\beta}$, results in an entanglemon qubit well protected from depolarization errors but susceptible to dephasing. The second, model $\mathbb{Z}_{2}^{\beta}$, results in entanglemon qubits protected from both kinds of errors. Then, we outline four different routes to possible realizations of such entanglemons in i) hyperfine states of trapped ions ii) Josephson junction arrays in superconducting circuits iii) quantum dots in graphene (or possibly silicon) and iv) quantum Hall skyrmions in graphene. Each platform has its own set of well-known strengths (regarding e.g.\ scalability, gate times, coherence times) and weaknesses, the most fundamental of which arguably is susceptibilty to noise. Hence, we find it very appealing that a general entanglemon construction from the heuristics of local entanglement can be concretely applied to obtain well-protected qubits in all these platforms.

The collective degree of freedom required for our entanglemon proposal arises from the construction of coherent states in $\mathbb{C}$P(3) space, and the corresponding quantum Hamiltonian, using Schwinger bosons. The basis states for the entanglemon qubit emerges as the lowest  energy doublet in the large Hilbert space dimension limit of the Schwinger bosons. In our hardware proposals we highlight how to obtain such a limit. Within the Schwinger boson Hilbert space (under a uniform noise approximation), the first model we present is exponentially (in system parameters) protected against depolarization while offering no protection from dephasing. Whereas, the second model offers exponential protection against both. Further, physical noise operators can induce higher-order matrix elements by pushing the system out of the Schwinger boson Hilbert space, forcing one to go beyond the uniform noise approximations. We show that for the second model, such noise decays exponentially in system parameters for depolarization and only appears in second order for dephasing.

\section{Background}
\label{sec_background}
Here, we provide a mainly pedagogical account of the basic principle underlying our proposed construction, before  delving into details in the subsequent sections. 
First, we introduce the basic idea of the entanglemon arising from the entanglement phase of a pair of degrees of freedom. Second, we provide a brief description of the general principle behind the entanglemon construction drawing parallels from more familiar constructions. Third, we present a heuristic explanation for the noise-protection afforded to the entanglemon qubit due to the emergence of an additional weakly coupled degree of freedom arising from the non-linearity of complex projective space.

\begin{figure*}
    \centering
    \includegraphics[scale = 0.37]{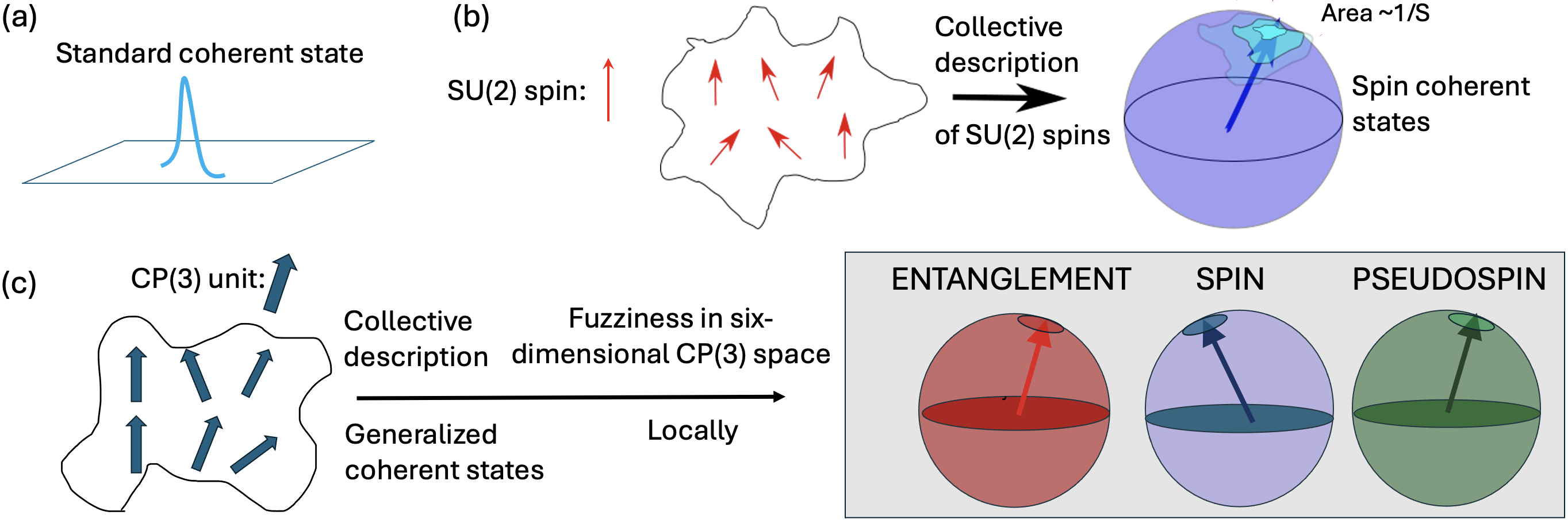}
    \caption{Pictorial description of the general principle behind the entanglemon construction. a) A pictorial representation of a standard minimum uncertainty coherent state of a harmonic oscillator with a peak around a well-defined value in phase space. b) The generalization of a coherent state to a collective state of many SU(2) spins. The phase space for the collective spin coherent state is $S^2$ and hence the spin coherent state can also be parametrized by the Bloch sphere, however, unlike the coherent state the spin coherent state does not peak at a fixed value. There is a "fuzziness" due to quantum fluctuations which vanishes in the large-S limit in which the spin coherent states reduce to the standard coherent states. c) Generalization of the spin coherent state construction to $\mathbb{C}$P(3) coherent states. The collective state of a group of $\mathbb{C}$P(3) objects has a six-dimensional phase-space corresponding to $\mathbb{C}$P(3). Similar to (b) there is an inherent fuzziness in the peak value which in the $d \rightarrow \infty$ limit corresponds to the standard coherent states (classical limit). Locally, the phase-space can be represented by three separate Bloch spheres via a Schmidt decomposition (see sec. \ref{subsec_nontech}) which gives rise to the notion of spin, pseudospin and entanglement Bloch spheres.}
     \label{fig1}
\end{figure*}

\subsection{Non-technical summary}
\label{subsec_nontech}
We start with an account of the concrete origin of our basic degree of freedom, the `entanglement phase' $\beta$. To do so, consider
the quantum state of an object with a spin-1/2 and a pseudospin-1/2 degree of freedom. Such a state can can be represented by a normalized 4-component complex spinor $\ket{\psi(\bm{r})}$. Since the global phase of this spinor is not physically observable, the phase space for such systems is $\mathbb{C}$P(3). Any such spinor can be written via a Schmidt decomposition in the form 
\begin{equation}
\begin{aligned}
    \ket{\Psi (\bf{r})} &= \rm{cos}\frac{\alpha}{2}\ket{\phi_S}\ket{\phi_P} + e^{i\beta}\rm{sin}\frac{\alpha}{2} \ket{\chi_S}\ket{\chi_P}\\
    \ket{\phi_{S(P)}} &= \bigg(\rm{cos}\frac{\theta_{S(P)}}{2},e^{i\varphi_{S(P)}}\rm{sin}\frac{\theta_{S(P)}}{2}\bigg)^T\\
    \quad  \ket{\chi_{S(P)}} &= \bigg(-e^{-i\varphi_{S(P)}}\rm{sin}\frac{\theta_{S(P)}}{2},\rm{cos}\frac{\theta_{S(P)}}{2}\bigg)^T
    \label{schmidtdec}
\end{aligned}
\end{equation}
where $\ket{\phi_{S(P)}}$,$\ket{\chi_{S(P)}}$ are orthogonal basis vectors for the spin (pseudospin) subspace. The $\mathbb{C}$P(3) phase space is parametrized by ($\theta_S,\varphi_S,\theta_P,\varphi_P,\alpha,\beta$), which pairwise parametrize three Bloch spheres corresponding to spin, pseudospin and entanglement (see Fig. \ref{fig1}) \cite{Doucotent}. Now if $\alpha = 0$ or $\pi$, the resulting state is a product state. However, for other values, the state is entangled. Recent work has shown novel symmetry broken ground states associated with anisotropic patterns of the entanglement measure, related to $\alpha$ \cite{chakrabortysmectic}. 

For the entanglemon, however, we are concerned with the entanglement phase $\beta$ and the entanglement Bloch sphere embedded in higher dimensional complex projective space. The entanglement phase, unlike the global phase, is a relative phase in a quantum superposition and is hence physical.

The central idea behind the entanglemon qubit is based on 
promoting this entanglement phase $\beta$ 
as the main degree of freedom. To do so, we propose a setting where $\beta$ 
 emerges as a collective variable for a set
 of $d$ copies of such pairs of spin-1/2 systems (general principle in the next subsection). We present two simple models for entanglemons in which the qubit is encoded: in  the lowest states of an anharmonic spectrum in the first, and for a double well potential in the second. In both cases, the qubit states belong to part of the spectrum associated with the collective coordinate $\beta$, an additional compact degree of freedom which arises due to non-linear geometry of $\mathbb{C}$P(3) phase-space. It is this feature which also endows the qubits with varying levels of immunity from noise. Hence, the entanglemon construction is a general and cross-platform construction for qubits protected from noise.

\subsection{General principle for entanglemon construction}
\label{sec_genprinc}
The entanglemon construction hinges on the emergence of quantum collective degrees of freedom in the low-energy susbspace of a collection of a pair of spin-1/2 degrees of freedom. In this subsection we explain how such collective degrees of freedom are expressed through generalized coherent states, giving rise to the notion of a quantum $\hat{\beta}$ degree of freedom. We shall also highlight how the number of pairs allows us to tune between quantum and classical limits. We only focus on the low-energy symmetric subspace in this section, to see how such a subspace is embedded in the full Hilbert space of a microscopic system refer to sec. \ref{sec_implementations} where we discuss physical implementations. In all such implementations, the least energetic states are the least energetic states of the symmetric subspace, for which the coherent state description developed in this section is valid.

Let us consider a more familiar and friendlier example first to understand the coherent state picture of the symmetric subspace. Take a collection of $2S$ ferromagnetically coupled SU(2) spins 1/2, resulting in a Hilbert space of dimension $2^{2S}$. The low energy subspace is symmetric (under permutations), its dimension is $\mathrm{dim}(\mathcal{H}_{sym}) = 2S+1$
and it is realized as the Hilbert space of Schwinger boson operators ($n_{\uparrow}$, $n_{\downarrow}$) with the constraint $n_{\uparrow} + n_{\downarrow} = 2S$.  Using these Schwinger bosons one can construct spin-$S$ coherent states which are parametrized by the Bloch sphere $S^2$ (also the Hilbert space of a single spin-1/2, modulo an overal phase factor ) \cite{radcliffe1971some}.

In this case, there is a uniform collective mode associated to spontaneously breaking the global SU(2) symmetry. Geometrically this can be thought of as the spins collectively pointing towards a direction represented by two angles which parameterize the Bloch sphere. However, there is an inherent fuzziness in this value due to quantum fluctuations which depend on $S$. As a result, instead of pointing in a specific value of the angles one can think of the spins pointing across a spread of values whose area in the Bloch sphere scales as $1/S$ (see Fig. \ref{fig1}b). The classical limit of these spin-coherent states is achieved by taking $S \rightarrow \infty$ at which point they become the standard coherent states for harmonic oscillators and are peaked along a specific direction in phase-space.

The emergence of collective degrees of freedom for the entanglemon follows a similar route, however, instead of a group of SU(2) spins we consider $d$ blocks of a pair of spin-1/2 degrees of freedom. The quantum Hilbert space attached to each block is four
dimensional, so it realizes a copy of the standard fundamental 
four-dimensional representation of SU(4). 
The dimension of the full Hilbert space with $d$ copies is $4^d$, and the corresponding SU(4) representation decomposes into a direct sum of irreducible representations. Among them is the fully symmetric subspace with respect to permutation between copies, which is the Hilbert space generated by the four Schwinger bosons with dimension $\mathrm{dim}(\mathcal{H}_{sym}) = (d+3)(d+2)(d+1)/6$. Similar to spin coherent states we can construct generalized $\mathbb{C}$P(3) (see sec. \ref{subsec_quantcoherent} and appendix \ref{subsec_geomquant}) coherent states which belong to this symmetric subspace and are expressed in terms of four Schwinger bosons (since we now have two spin-1/2s per unit). In this general introduction we shall not discuss how to implement a model whose low-energy subspace is the symmetric subspace, see sec \ref{subsec_impl_Mod_1} for a discussion on that. 

The generalized $\mathbb{C}$P(3) coherent states belong to a six dimensional phase space that can be represented locally by three Bloch spheres. For spin coherent states, the collective mode could be represented by a fuzzy sphere (see Fig. \ref{fig1}b). Similarly, for the entanglemon we get three collective modes, one of which is represented by the entanglement Bloch sphere (see sec. \ref{subsec_nontech} and Fig. \ref{fig1}) parameterized now by quantum degrees of freedom, $\hat{\alpha}$ and the entanglement phase $\hat{\beta}$. 

Once again, like the spin coherent states, the connection between quantum and classical is provided by the parameter $d$ which tunes the fuzziness induced by quantum fluctuations. In the $d \rightarrow \infty$ limit, the
generalized coherent states too are peaked around well-defined points on the six dimensional
$\mathbb{C}$P(3) manifold. However, for finite $d$, quantum fluctuations of single boson observables around one such point are proportional to $1/d$, so this quantity plays the role
of an effective Planck's constant and
allows us to interpolate between the "most" quantum case for $d=1$
and the classical limit $d \rightarrow \infty$. The $d=1$ case is special in the
sense that all quantum states turn out to be coherent states, because $\mathbb{C}$P(3) is
isomorphic to the fundamental four-dimensional representation of SU(4) modulo an overall
phase factor that is physically undetectable.

Another familiar collective mode quantization story is that of superconducting qubits (or the quantum XY model). For the case of superconductors, the story starts with Leggett's observation that collective modes of superconducting circuits, resulting from the spontaneously broken global U(1) symmetry,
can behave quantum mechanically. Such collective modes correspond to phase differences across Josephson junctions. This is the quantum degree of freedom that is used in  
modern superconducting qubits such as the transmon. The underlying classical phase-space
is then the cylinder $S^1 \times \mathbb{R}$ with coordinates $\varphi$ (phase of the superconducting order parameter) and $L^z$ (z-component of angular momentum for quantum rotors) or $N$ (number of cooper pairs for superconductors). Importantly, here the phase-space is non-compact and depends on physical parameters of the microscopic model, eg: $E_J/E_C$ for a Josephson junction.

While the collective mode quantization for the entanglemon is similar to these above constructions, there is a differentiating component. Such a difference comes due to the higher-dimensional complex projective space in which the entanglement Bloch sphere is embedded and is responsible for the noise immunity of the qubit. The entanglemon branch of the spectrum is accompanied by two other harmonic oscillator modes corresponding to standard collective oscillations of spins and pseudospins, respectively. However, in the models we present, the two least energetic states of the entanglemon spectrum are the lowest energy states of the model, 
and are well separated from all the other states. The entanglemon spectrum, for model $U(1)^{\beta}$ has an anharmonic spectrum (like in the transmon, although with important differences). For the other model, model $\mathbb{Z}^\beta_2$, the continuous symmetry is reduced to the discrete $\mathbb{Z}_2$ subgroup
and the qubit states are the associated doublet states.

Importantly, note that the entanglemon does not rely on any platform specific details such as superconductivity, ion structure, etc. The construction simply relies on there being two spin-1/2 degree of freedom on each "site", the possibility of symmetric low-energy states with entanglement between the two and the quantization of the corresponding entanglement phase.

\subsection{Physical heuristic for noise protection}

The main reason for the protection of the entanglemon qubit, is the presence of the additional $\hat{\beta}$ collective quantum degree of freedom arising due to the non-linear geometry of higher-dimensional complex projective space. Local and physical noise operators are insensitive to translations in $\hat{\beta}$ (since the simplest noise terms don't depend on such a collective correlation), hence encoding information in the spectrum associated with the $\beta$-direction in six-dimensional complex projective space, $\mathbb{C}$P(3), allows a high-degree of protection. 

We discuss the detailed application of the above heuristic and the resulting levels of noise protection for the two simple models we present in this work, in sections \ref{sub_noise_protection_dep}, \ref{subsec_full_protection} and \ref{sec_noiseanalysis}. In both  models, the reason for protection against depolarization will be disjoint support, i.e distinct eigenvalues of the operator $\hat{G}$ which generates the symmetry associated with $\hat{\beta}$. For the second model, the additional protection against dephasing will come from almost discrete values of $\beta$ forming the basis for the logical qubit Hilbert space. This results in a broad spread in the conjugate variable, thus rendering the two qubit states indistinguishable in measurements of $\hat{G}$.

\section{simple models for entanglemon qubits}
\label{sec_simplemodels}
In this section we present two simple models for entanglemon qubits. These models describe the symmetric subspace to which the entanglemon qubit states belong. To see how such a subspace is embedded in a much larger physical Hilbert space refer to sec. \ref{sec_implementations}. Both these models are completely platform-agnostic. The first model is protected from depolarization errors whereas the second is protected from both depolarization and dephasing errors. We present these two simple models in full detail - including the quantization, the different modes corresponding to the qubit states and their respective noise immunities. Readers interested in more platform specific constructions can read this section in combination with section V, where we discuss the multitude of platforms where such noise-robust qubits could be constructed.

\begin{figure*}
    \centering
    \includegraphics[scale = 0.5]{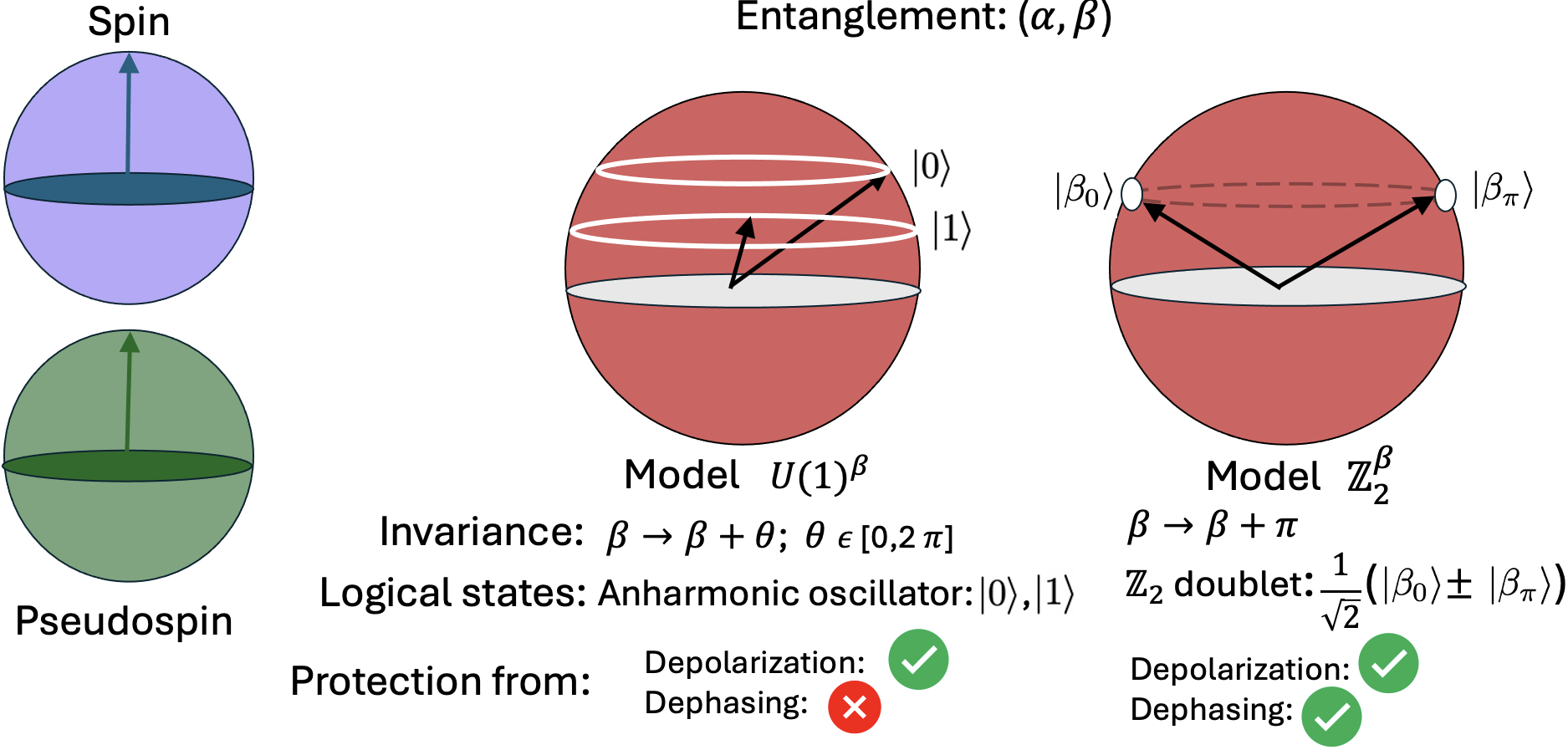}
    \caption{Pictorial description of the two simple models for entanglemons --- models $U(1)^\beta$ and $\mathbb{Z}^\beta_2$ --- in terms of the three Bloch spheres, spin (violet), pseudospin (green) and entanglement (red), which parametrize the generalized $\mathbb{C}$P(3) coherent state (see Fig. \ref{fig1}). Both models involve entanglement between spin and pseudospin degrees of freedom where the qubit states are easy-axis states, i,e spin and pseudospin point along the z-axis. For model $U(1)^{\beta}$, the quantum states corresponding to the qubit are the orbits in the entanglement Bloch sphere corresponding to different values of $\alpha$. These states have a continuous $U(1)^{\beta}$ symmetry corrsponding to the transformation $\beta \rightarrow \beta + \theta$; $\theta \in [0,2\pi)$. These two states represent the two lowest states of an anharmonic oscillator spectrum which are well separated from the higher energy states (due to anharmonicity) and from other oscillator modes of the $\mathbb{C}$P(3) system (see sec III. 4. and Fig. \ref{specfig}). Such an encoding gives the qubit excellent protection noise-induced depolarization errors, however, there is no protection from dephasing (see sec \ref{sub_noise_protection_dep}). Model $\mathbb{Z}^\beta_2$ represents a different encoding for the entanglemon, where the continuous $U(1)^\beta$ of the previous model is broken down to a discrete $\mathbb{Z}_2^\beta$ symmetry. The two states comprising the basis for the logical qubit Hilbert space are then superpositions of states which lie on the same $\alpha$ orbit but at opposite $\beta$ points. Explicitly written, the qubit states are $\ket{0} = (\ket{\beta_0}+\ket{\beta_\pi})/\sqrt{2}$ and $\ket{1} = (\ket{\beta_0}-\ket{\beta_\pi})/\sqrt{2}$ and the qubit is protected from both depolarization and dephasing errors (see sec \ref{subsec_full_protection}).}
    \label{figmodels}
\end{figure*}

\subsection{Model $U(1)^{\beta}$: Integrable entanglemon model}
We start with a completely classical description, following which we quantize the system to obtain the entanglemon spectrum. We then illustrate how using the lowest doublet of entanglemon states results in a qubit immune to depolarization errors. However, such a construction is still suscpetible to dephasing, we outline a general strategy to tackle such dephasing at the end of subsection A.

\subsubsection{Classical Description}
\label{subsec_class_desc}
We start with a classical energy functional (Hamiltonian)
\begin{equation}
    H_C = u_p (P_x^2+P_y^2) + u_z P_z^2 - S_z
    \label{classH}
\end{equation}
where $u_p,u_z \geq 0$ and the
spin and pseudospin components are functions defined on the
$\mathbb{C}$P$^3$ manifold, according to:
\begin{equation}
A(\bm{\Psi}) = \frac{\bm{\Psi}^{\dagger}\hat{A}\bm{\Psi}}{\bm{\Psi}^{\dagger}\bm{\Psi}} = \langle \hat{A} \rangle_{\bm{\Psi}} 
\label{optoexp}
\end{equation}
where $\bm{\Psi}$ belongs to the four dimensional complex space associated
to the spin and pseudospin, and $\hat{A}$ is any operator ($S$ or $P$) acting on this space.
$A(\bm{\Psi})$ is unchanged as $\bm{\Psi}$ is multiplied by an arbitrary
non vanishing complex number, hence, it can be used to define a function on the
projective space $\mathbb{C}$P$^3$.

The ground states of the above system exhibit entanglement in the sense described in sec IIA \cite{LianPRL}. These entangled ground states can be split into two classes: i) Easy-axis for $u_p > u_z > 1/2$
\begin{equation}
    \ket{\psi(\bm{r})} = \big(\text{cos}\frac{\alpha^*}{2},0,0,e^{i\beta}\text{sin}\frac{\alpha^*}{2}\big)^T,\big(0,\text{cos}\frac{\alpha^*}{2},e^{i\beta}\text{sin}\frac{\alpha^*}{2},0\big)^T
    \label{eags}
\end{equation}
and ii) easy-plane for $u_z > u_p > 1/2$ 
\begin{equation}
\begin{split}
    \ket{\psi(\bm{r})} = \bigg(\text{cos}\frac{\alpha^*}{2},-e^{i(\beta-\varphi_P)}\text{sin}\frac{\alpha^*}{2},
    e^{i\beta}\text{cos}\frac{\alpha^*}{2},e^{i\beta}\text{sin}\frac{\alpha^*}{2}\bigg)^T
\end{split}
\label{epgs}
\end{equation}
where $\alpha^* = \text{sec}^{-1}(2u_z)$, the x,y and z components of the psin and pseudospin are derived from eq. \ref{optoexp} and easy-axis/plane corresponds to whether the pseudspin degree of freedom lies on the x-y plane or along the z-axis. In both cases, we have a residual $U(1)^{\beta}$ symmetry corresponding to $\beta \rightarrow \beta + \theta$; for all $\theta \in (0,2 \pi)$.

For the quantization, the following features that occur at the classical level are important. Besides the $U(1)^{\beta}$ that is present in the ground state Hamiltonian, there are two explicit U(1) symmetries: $U(1)_S$ and $U(1)_P$ corresponding to the rotations of the spin and pseudospin Bloch vectors about their z-axis. Quantizing the system will result in a discrete spectrum with these different kinds of modes (see Fig. \ref{sec_spectrum}). Second, there is a Poisson bracket structure associated with the parameters $\alpha,\beta$ in this six-dimensional phase-space ($\theta_S,\theta_P.\varphi_S,\varphi_P, \alpha,\beta$), resulting in a notion of Hamiltonian flow.

\subsubsection{Quantization via Schwinger bosons and generalized coherent states}
\label{subsec_quantcoherent}
To quantize the system described above we introduce four Schwinger boson operators, one for each component of the complex spinor, i.e $\psi_j \rightarrow \hat{a}_j$ for $j = 1,2,3,4$. The physical Hilbert space of such a mapping is defined by the constraint
\begin{equation}
   n_1 + n_2 +n_3 +n_4 = d
   \label{eq_constraint}
\end{equation}
where $n_j =\langle \hat{n}_j \rangle = \langle \hat{a}^{\dagger}_j \hat{a}_j \rangle$ is the expectation value of $j$ bosons. The physical Hilbert space arising from the above constraint has dimension $(d+3)(d+2)(d+1)/6$. 

Further from eq. \ref{optoexp} and $\psi_j \rightarrow a_j$ one obtains the following expressions for spin and pseudospin operators in terms of the Schwinger bosons
\begin{equation}
\begin{split}
    \hat{S}_x &= (\hat{a}^{\dagger}_1\hat{a}_2 + \hat{a}^{\dagger}_2 \hat{a}_1+ \hat{a}^{\dagger}_3\hat{a}_4+ \hat{a}^{\dagger}_4\hat{a}_3)/d \\
     \hat{S}_y &= i(\hat{a}^{\dagger}_2\hat{a}_1 - \hat{a}^{\dagger}_1 \hat{a}_2+ \hat{a}^{\dagger}_4\hat{a}_3 - \hat{a}^{\dagger}_3\hat{a}_4)/d \\
     \hat{S}_z &= (\hat{a}^{\dagger}_1\hat{a}_1 - \hat{a}^{\dagger}_2 \hat{a}_2+ \hat{a}^{\dagger}_3\hat{a}_3 - \hat{a}^{\dagger}_4\hat{a}_4)/d \\
     \hat{P}_x &= (\hat{a}^{\dagger}_1\hat{a}_3 + \hat{a}^{\dagger}_3 \hat{a}_1+ \hat{a}^{\dagger}_2\hat{a}_4+ \hat{a}^{\dagger}_4\hat{a}_2)/d \\
     \hat{P}_y &= i(\hat{a}^{\dagger}_3\hat{a}_1 - \hat{a}^{\dagger}_1 \hat{a}_3+ \hat{a}^{\dagger}_4\hat{a}_2 - \hat{a}^{\dagger}_2\hat{a}_4)/d \\
     \hat{P}_z &= (\hat{a}^{\dagger}_1\hat{a}_1 + \hat{a}^{\dagger}_2 \hat{a}_2- \hat{a}^{\dagger}_3\hat{a}_3 - \hat{a}^{\dagger}_4\hat{a}_4)/d .\\
     \label{sbosrep}
\end{split}
\end{equation}
Using the above operators, one can express the quantum version of $H_C$ (eq. \ref{classH}), in each sector labelled by $d$. As mentioned in earlier sections, the classical limit is then the $d \rightarrow \infty$ limit. The quantum Hamiltonian is
\begin{equation}
\begin{split}
    \hat{H}_{Qd} = \frac{4u_{p}}{d^2}(\hat{n}_1 \hat{n}_2 + \hat{n}_3 \hat{n}_4 + \hat{a}^{\dagger}_1\hat{a}_2\hat{a}_3\hat{a}^{\dagger}_4 + \hat{a}_1\hat{a}_2^{\dagger}\hat{a}_3^{\dagger}\hat{a}_4) + \\
    \frac{u_z}{d^2}(\hat{n}_1 - \hat{n}_2 + \hat{n}_3 - \hat{n}_4)^2 -
    \frac{1}{d}(\hat{n}_1 + \hat{n}_2 - \hat{n}_3 - \hat{n_4}) + \text{const}
    \label{hqham}
\end{split}
\end{equation}
To connect the classical and quantum regimes we implement a generalized coherent state construction for $\mathbb{C}$P(3) and a geometric quantization procedure using such coherent states (see appendix \ref{subsec_geomquant} for details). For $\bm{\Psi} \in \mathbb{C}^4$ we define $\hat{b}^{\dagger}(\bm{\Psi}) = \sum_{j=1}^{4}\Psi_j \hat{a}_j^{\dagger}/\sqrt{\bm{\Psi}^{\dagger}\bm{\Psi}}$ so that $[\hat{b},\hat{b}^{\dagger}] = 1$. We can the define the coherent state
\begin{equation}
    \ket{CS_d(\bm{\Psi})} \equiv \frac{1}{\sqrt{d!}} \hat{b}^{\dagger ^d}(\bm{\Psi}) \ket{0}
\end{equation}
which is normalized, $\langle CS_d(\bm{\Psi})|CS_d(\bm{\Psi})\rangle = 1$. The above construction gives the useful relation
\begin{equation}
    \langle CS_d(\bm{\Psi})|a^{\dagger}_i a_j | CS_d(\bm{\Psi}) \rangle = \frac{d \Psi_i^* \Psi_j}{\bm{\Psi}^{\dagger}\bm{\Psi}}
    \label{expvalue}
\end{equation}
which can be obtained from $\hat{a}_j \hat{b}^{\dagger ^d}(\bm{\Psi})\ket{0} = [\hat{a}_j,\hat{b}^{\dagger ^d} (\bm{\Psi})]\ket{0} = [\hat{a}_j,\hat{b}^{\dagger} (\bm{\Psi})]d \hat{b}^{\dagger ^{d-1}} (\bm{\Psi})\ket{0} = d \frac{\Psi_j}
{\sqrt{\bm{\Psi}^{\dagger}\bm{\Psi}}} \hat{b}^{\dagger ^{d-1}} (\bm{\Psi})\ket{0}$. 
From the above relations, if we denote the quantum Hamiltonian obtained from the Schwinger bosons by $\hat{H_Q}$,  we obtain
\begin{equation}
    \langle CS_d(\bm{\Psi}) | \hat{H}_{Qd} | CS_d(\bm{\Psi}) \rangle = H(\bm{\Psi})
    \label{def_classical_energy_functional}
\end{equation}

In the language of geometric quantization, the above relation implies that the classical energy function $H(\bm{\Psi})$ over $\mathbb{C}$P(3) is the covariant symbol of the quantum Hamiltonian $\hat{H}_Q$. In the large
$d$ limit, the function $H(\bm{\Psi})$ coincides with the classical energy
given by Eq.~(\ref{classH}).

\subsubsection{Quantum Integrability}
\label{subsec_integ}
Before analyzing the spectrum of $H_{Qd}$, we can obtain a great deal of physical insight by exploiting the simplicity of our model. Note that, besides the total energy, the Schwinger boson Hamiltonian $H_Q$ has two other conserved quantities: $n_1 - n_4$ and $n_2-n_3$. Hence, with three degrees of freedom and three conserved quantities we have an integrable quantum system. 

Here, we present a geometrical description which allows us to infer the nature of the various modes in our system. Remember that the set of allowed states is defined by the constant $n_1 + n_2 + n_3 + n_4 = d$. Such a constraint defines the interior of a tetrahedron in $\mathbb{N}^4$. We can visualize this tetrahedron using the basis variables $(n_1-n_4,n_2-n_3,n_1+n_4)$. The four vertices of this tetrahedron are given by $n_1 = d \rightarrow (d,0,d)$, $n_2 = d \rightarrow (0,d,0)$, $n_3 = d \rightarrow (0,-d,0)$ and $n_4 = d \rightarrow (-d,0,d)$.

Let us first consider the easy-axis case, with $u_p > u_z > 1/2$. We know from the previous section that for this case there are two ground states as in eq. \ref{eags}. These ground states in the $(n_1-n_4,n_2-n_3,n_1+n_4)$ coordinates are $\text{CGS}_1 \equiv (d\, \cos{\alpha^*},0,d)$ and $\text{CGS}_2 \equiv (0,-d\, \cos{\alpha^*},0)$. Since $\cos{\alpha^*} \neq 0$, states in the neighborhood of $\text{CGS}_1$ are exactly degenerate with states in the neighborhood of $\rm{CGS}_2$ because their quantum numbers ($n_1-n_4$,$n_2-n_3$) are different. In other words, this twofold degeneracy is protected by symmetry (Ising like) and hence no tunneling induced splittings occur.

We now consider the excitations near these ground states. The entanglemon excitations for the easy-axis case, corresponds to moving along the edges of the tetrahedron. About $\text{CGS}_1$, the entanglemon corresponds to varying $n_1-n_4$ at fixed $n_2-n_3 = 0$ and the oscillator modes correspond to changing both these quantum numbers (Note that different signs of $n_2-n_3$ correspond to different faces of the tetrahedron and hence to different excitations).

In the easy-plane case with $u_z > u_p > 1/2$,  the ground state now lies \textit{inside} the tetrahedron, with coordinates ($\frac{d}{2}\, \text{cos}\alpha^*,-\frac{d}{2}\, \text{cos}\alpha^*,\frac{d}{2}$). Since for the rest of this section we shall focus on the easy-axis case, we do not present the figure for the geometrical analysis of the easy-plane case here. However, one can expect that in this case, the entanglemon mode  corresponds to moving across the line $n_1-n_4 = n_2-n_3$.
\begin{figure}
    \centering
    \includegraphics[scale = 0.45]{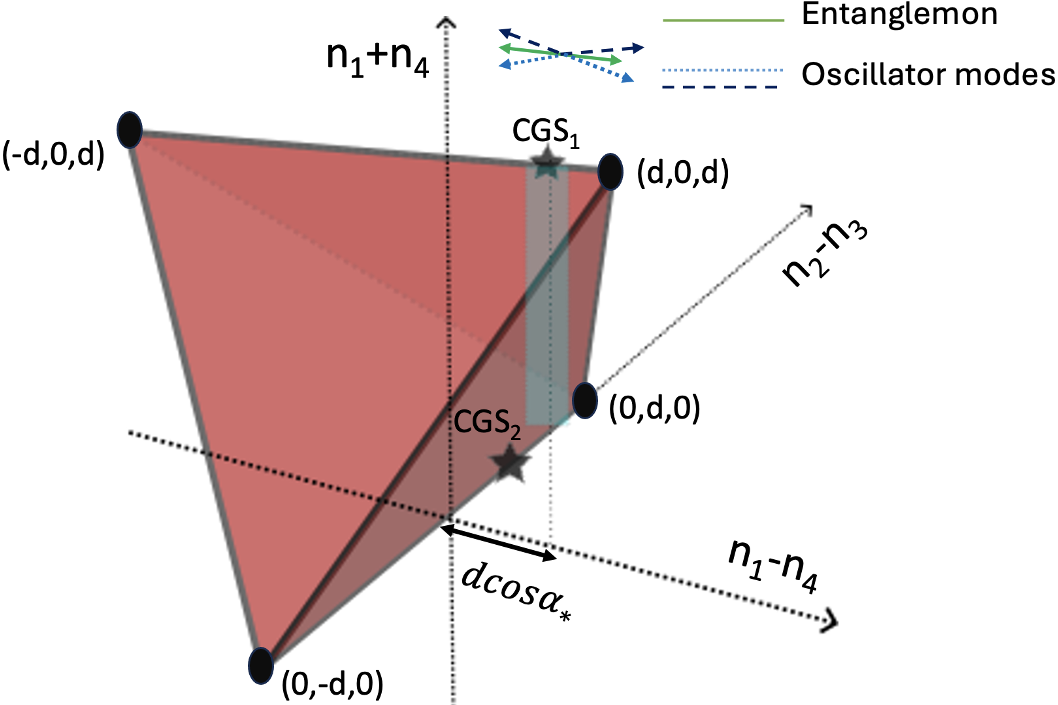}
    \caption{Geometry of Quantum Integrability - Tetrahedron of possible states with the two possible ground states for the easy-axis case marked $\text{CGS}_1$ and $\text{CGS}_2$. The entanglemon excitation about $\text{CGS}_1$ corresponds to moving along the top edge whereas the other oscillator modes correspond to moving along the faces. The former has an anharmonic spectrum associated to the compact variable $\beta$, whereas the latter oscillator modes have the standard harmonic oscillator spectrum. The excitation spectra in the neighbourhood of $CGS_1$ along the grey rod is computed analytically (semi-classical large-$d$ approximation) in sec. \ref{sec_spectrum}}
    \label{Eaxinteg}
\end{figure}

\subsubsection{Entanglemon spectrum}
\label{sec_spectrum}
    
We now calculate the spectrum of the quantum Hamiltonian $\hat{H}_{Qd}$ given in eq. \ref{hqham}. Since in the vicinity of $\text{CGS}_\text{1}$ (see eqs. \ref{eags} and \ref{expvalue} ) in  fig. \ref{Eaxinteg}, $n_2$ and $n_3$ are small and, in the large-d (semiclassical) limit, $n_1$ and $n_4$ are large, we can reduce the quartic $\hat{H}_Q$ to a quadratic one by assuming $\hat{a}_1,\hat{a}^{\dagger}_1 = \sqrt{n_1}$ and $\hat{a}_4,\hat{a}^{\dagger}_4 = \sqrt{n_4}$. 

Note that replacing the quartic
$\hat{a}_1\hat{a}_2^{\dagger}\hat{a}_3^{\dagger}\hat{a}_4$ 
operator by the simpler quadratic approximation
$\sqrt{n_1 n_4}\,\hat{a}_2^{\dagger}\hat{a}_3^{\dagger}$ violates the
constraint on the total number of Schwinger bosons, since $n_1$
and $n_4$ are no longer treated as operators but as classical parameters.
This is justified in the large $d$ limit where quantum fluctuations
of these quantities are small. But in view to storing quantum information, it is crucial to keep a correct counting of low energy quantum states.
The constraint Eq.~(\ref{eq_constraint}) implies that
the "classical parameters" $n_1$ and $n_4$ should be chosen as integers
such that the parity of $n_1-n_4$ should be equal to the parity of
$d + \hat{n}_2 - \hat{n}_3$. Note that the later quantity is well defined,
because it commutes with the approximate quadratic Hamiltonian as given
below.

This quadratic approximation gives us 
\begin{equation}
\begin{split}
    \hat{H}_{Qd} \approx \frac{4u_p}{d^2}(n_1 \hat{a}^{\dagger}_2\hat{a}_2 + n_4 \hat{a}^{\dagger}_3\hat{a}_3) + \frac{4u_p \sqrt{n_1 n_4}}{d^2}(\hat{a}^{\dagger}_2 \hat{a}^{\dagger}_3 + \hat{a}_2 \hat{a}_3)\\
    + \frac{u_z}{d^2}(n_1-\hat{a}^{\dagger}_2\hat{a}_2+\hat{a}^{\dagger}_3\hat{a}_3-n_4)^2 -\frac{1}{d}(n_1-n_4+\hat{a}^{\dagger}_2\hat{a}_2-\hat{a}^{\dagger}_3\hat{a}_3)\\
    +2\frac{u_p}{d}\\
    \end{split}
\end{equation}
Expanding the first term in the second line of the above equation, using that $n_2-n_3$ is small (see eq. \ref{eags} and Fig. \ref{Eaxinteg}) we get a quadratic Hamiltonian in terms of the Schwinger boson operators $\hat{a}_2, a_2^{\dagger}$ and $\hat{a}_3, a_3^{\dagger}$. 
We diagonalise this using the  Bogoliubov transformation
\begin{equation}
    \hat{a}^{\dagger}_2 = l\hat{x}^{\dagger} + m\hat{y}\ \  ; \ \
    \hat{a}^{\dagger}_3 = l\hat{y}^{\dagger} + m\hat{x}
\end{equation}
with $x$ and $y$ chosen so that the off-diagonal terms vanish:
\begin{align}
\begin{split}
    \hat{H}_{Qd} = E_T &+ \omega_x(\hat{x}^{\dagger}\hat{x} + \frac{1}{2}) + \omega_y(\hat{y}^{\dagger}\hat{y} + \frac{1}{2});\\
    \omega_x &= \frac{1}{d^2}(4u_p-2u_z)(n_1-n_4)-\frac{1}{d};\\
    \omega_y  &= \frac{1}{d^2}2u_z(n_1-n_4) + \frac{1}{d};\\
    E_T = &\dfrac{u_z (n_1-n_4)^2}{d^2} - \dfrac{(n_1-n_4)}{d} \, .
    \label{spec}
\end{split}
\end{align}
The above expression yields
three different classes of modes: i) An oscillator mode arising from $\hat{x}^{\dagger}\hat{x}$ term. Acting with $x^{\dagger}$ increases $n_2-n_3$ by 1 and $\omega_x = \omega_x (n_1-n_4)$. ii) Another oscillator mode, which corresponds to moving along the dotted line in Fig. \ref{Eaxinteg}, similar to the previous one in the sense that $\omega_y = \omega_y (n_1-n_4)$ but different since acting with $y^{\dagger}$ decreases $n_2-n_3$ by 1. Hence, for states around the rod near $\rm{CGS}_1$, i.e $n_2-n_3 = 0$, these two oscillator-like excitations correspond to moving along the opposite faces of the tetrahedron (see caption and fig. \ref{Eaxinteg}). iii) Finally, arising from the $E_T$ term on considering quantized values of the quantum number $n_1-n_4$, we have the entanglemon mode. In Fig. \ref{Eaxinteg}, as mentioned earlier, this mode corresponds to moving along the edge (green arrows) for quantized values of $n_1-n_4$ at fixed $n_2-n_3 = 0$. We can see from the expression of $E_T$ in eq. (\ref{spec}), these energy levels resemble that of a transmon qubit for a superconductor with increasing energy spacing between consecutive energy levels.

Note that the parity of $\hat{n}_2 - \hat{n}_3$ is the same as the parity of $\hat{x}^{\dagger}\hat{x} \pm \hat{y}^{\dagger}\hat{y}$. The $n_1$ and $n_4$ integers should therefore be chosen so that the parity of $n_1-n_4$ equals  the parity of $d+\hat{x}^{\dagger}\hat{x} \pm \hat{y}^{\dagger}\hat{y}$. From the expression of $E_T$ in eq. (\ref{spec}), we see that the entanglemon has a parabolic dispersion, the minimum of which lies at $n_1-n_4 = d/(2u_z)$. The parabolic nature of the entanglemon mode is characteristic of a free-particle like collective mode spectrum of an approximate symmetry. Since $\beta$ is compact (periodic), its conjugate degree of freedom $\alpha$ is quantized and the corresponding spectrum resembles that of a particle on a circle which is parabolic, similar to the collective mode originating from moving around the circular minima in a mexican hat potential. Whereas, the other two oscillator modes (see sec. \ref{subsec_integ} for geometric description of the three different modes) have a harmonic oscillator like spectrum, and are similar to the modes along the radial direction of a mexican hat potential (Higgs modes). Note, that it is the geometric non-linearity of $\mathbb{C}$P(3) space that allows for the presence of this approximate symmetry and the additional collective mode with the anharmonic spectrum. 

Even though the entanglemon states are at integer values of $n_1-n_4$, the minimum of this parabola at $d/(2u_z)$ is generically not an integer. Such a scenario signifies the presence of a Berry-phase associated with the cyclic rotation of the phase $\beta$ from $0$ to $2\pi$ and results in a unique low-energy doublet which forms the qubit states. For more details about Berry-phase considerations in these settings we refer the reader to the appendix. Moreover, higher energy levels are further separated due  to the anharmonicity in the spectrum. Hence, the energy difference between the lowest two states is unique and these two states form the doublet of states ($\ket{0},\ket{1}$) of our entanglemon qubit.

\begin{figure}
    \centering
    \includegraphics[scale = 0.4]{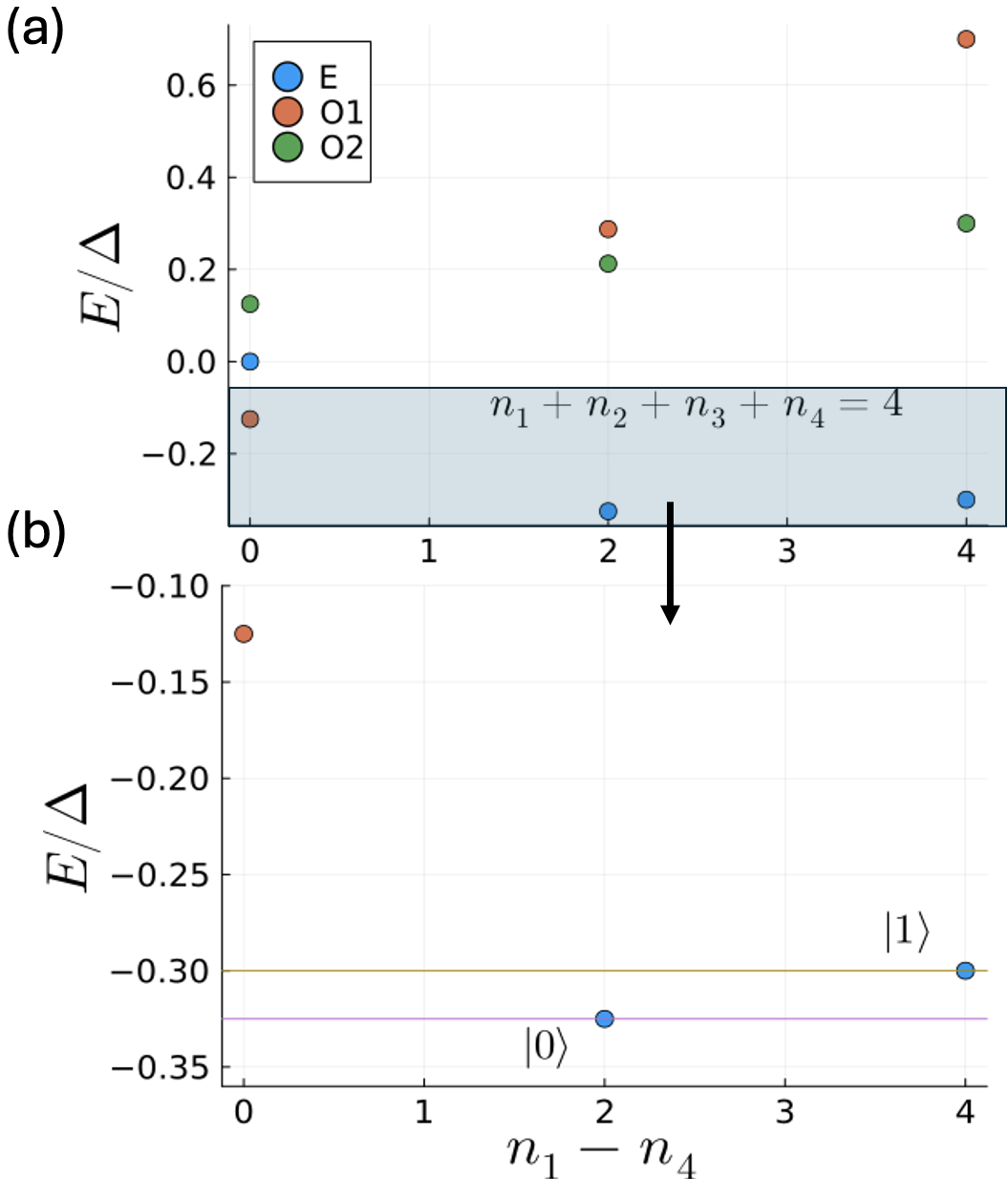}
    \caption{The low-lying spectrum as obtained in eq. (\ref{spec}) for $u_p = 2 \Delta$, $u_z = 0.7 \Delta$ and $d = 4$ and $n_2 - n_3 = 0$. (a)Blue-Entanglemon modes with anharmonic spectrum, Red and Green - Oscillator modes with standard harmonic spectrum. b) The lowest two modes of the anharmonic spectrum for the entanglemon qubit states $\ket{0}$ and $\ket{1}$ in model $U(1)^{\beta}$. The energy difference between these two states is  unique allowing for selective operations. The minima of the curve for the spectrum is generically at a non-integer value  (between 2 and 3 for these parameters), representing the Berry phase contribution from a cyclic $2\pi$ rotation of $\beta$ (see text for details.)}
    \label{specfig}
\end{figure}

\subsubsection{Noise Protection - Depolarization immunity}
\label{sub_noise_protection_dep}

For the entanglemon, we consider noise terms are described by operators of the form $\sum_i \hat{A}_i \otimes \hat{B}_i$, where $i$ labels one of the $d$ sites according to the general setting of section II. $\hat{A}_i$ acts in the local
Hilbert space (combination of spin 1/2 and pseudospin 1/2) at site $i$, and $\hat{B}_i$ acts in the much larger Hilbert space of the environmental degrees of freedom that cause relaxation and/or decoherence. Let's first examine the effects (perturbative) of such noise terms for the $U(1)_\beta$ entanglemon systematically by considering symmetric noise terms which are described by $\hat{S} (\hat{P})$ operators in the Schwinger boson subspace described in the model above. For such noise $\hat{B}_i$ is independent of $i$. We discuss possible sources of such noise for different hardware platforms in sec VI, after proposing the respective implementations. We also postpone a more detailed noise protection analysis in which we consider local asymmetric noise effects as well, to sec VI.

Bit-flip errors can occur if  such noise operators have non-vanishing matrix elements with the two qubit states.
Further such noise is particularly harmful for standard qubit formulations, since in some cases they can induce correlated errors which make quantum error correction unfeasible.
Let us study the decoherence caused due to such noise by examining the matrix elements of the $\hat{S}_x$ (or any other component) operator with the $\ket{0},\ket{1}$ states. A useful piece of physical intuition is provided by looking at the classical $d \rightarrow \infty$ limit. In appendix ~\ref{App_sec_beta_translations}, we show that the
residual $U(1)^\beta$ symmetry is generated by the function
$G(\alpha,\beta,\theta_S,\theta_P,\varphi_S,\varphi_P)=-\frac{1}{2}\cos{\alpha}$.
Let us consider the matrix elements
of $\hat{S}_{x,y,z}$ and $\hat{P}_{x,y,z}$ operators, 
between these two entanglemon qubit states. Recall that these are expressed
in terms of bilinear combinations of elementary Schwinger boson operators
in Eq.~(\ref{sbosrep}). The key idea here, on which the whole intuition
of the entanglemon qubit was based, is that
these operators correspond (in the large $d$ limit) to functions over $\mathbb{C}$P(3) which are {\em invariant} under $\beta$ translations.
This implies that their Poisson brackets with $G$ vanish.

At the quantum level, this suggests
that matrix elements of $\hat{S}_{x,y,z}$ and $\hat{P}_{x,y,z}$
between two eigenstates with different eigenvalues of the quantum operator $\hat{G}$ (associated to the classical function $G$) also vanish.
Because the classical energy function $H(\bm{\Psi})$ is also
invariant under $\beta$ translations, we expect $\hat{H}$ to commute
with the quantum operator $\hat{G}$ associated to the classical
function $G$. Which would imply that the $\ket{0}$ and $\ket{1}$ states are
eigenvectors of $\hat{G}$ with distinct eigenvalues, as it is typical
for a collective excitation branch associated to a nearly broken global symmetry. This leads to the expectation that matrix elements of
$\hat{S}_{x,y,z}$ and $\hat{P}_{x,y,z}$ operators between the two
entanglemon qubit states vanish.

The heuristic above is based on the approximate correspondence between quantum mechanical operators and functions over the $\mathbb{C}$P(3) manifold. Therefore, it is important to test this conjecture by examining
the action of $\hat{S}_{x,y,z}$ and $\hat{P}_{x,y,z}$ operators
on the actual $\ket{0}$, and $\ket{1}$ quantum mechanical states.
As we saw in the previous section and in Fig. \ref{specfig},  $(n_1-n_4)_{\ket{1}} - (n_1 - n_4)_{\ket{0}} = 2$. The only operator that would have a finite matrix element between $\ket{0}$ and $\ket{1}$ would be of the form $\hat{a}^{\dagger}_4 a_1$ or $\hat{a}^{\dagger}_1 a_4$, however, as we see from eq. \ref{sbosrep}, no such term is present and hence $\langle 0|S(P)_{x(y,z)}|1 \rangle = 0$ and the qubit is immune to bit-flip errors (depolarization) up to \textit{first order} in such symmetric noise terms. Second order symmetric noise terms will induce transitions between the two logical states since they will comprise the terms described above. 

While the $U(1)_\beta$ entanglemon provides considerable protection against depolarization, this construction offers no protection from dephasing. Phase errors are not protected in the $U(1)_\beta$ entanglemon since the generator $G$
of the $\rm{U}(1)_\beta$ symmetry is a simple function of the degree of entanglement between the spin and pseudospin subsystems. And this implies that, if this degree changes, the lengths of the classical vectors 
$\langle \bm{S} \rangle_{\bm{\Psi}}$ and $\langle \bm{P} \rangle_{\bm{\Psi}}$
also change, together with at least some of their components. At the quantum
level, we expect then that
diagonal matrix elements of the noise operators are not equal, i.e $\langle 0|\hat{A} | 0 \rangle \neq \langle 1|\hat{A} | 1 \rangle$, whenever 
$\hat{A}$ is of the form $\hat{S}_l$ or $\hat{P}_l$ with $l=x,y,z$. As we see in Fig. \ref{figmodels}, the qubit states correspond to different orbits in the Bloch sphere with different values of $\alpha$. This makes the two qubit states "distinguishable" since $\alpha$, the conjugate variable, couples to the physical spin and pseudospin.

\subsubsection{A general principle to overcome the dephasing dilemma}
How does one make the qubit states with disjoint support (due to vanishing matrix elements as shown above) also indistinguishable? One may ask if dephasing errors due to such symmetric noise are unavoidable in such entanglemon constructions. Any entanglemon formed from a $U(1)^\beta$ symmetric system will have no protection against dephasing due to the reasons mentioned earlier. However, if one could reduce this $U(1)^\beta$ to a discrete symmetry $\mathbb{Z}_{2\beta}$ then such protection could be achieved. 
We present a pathway to do this symmetry reduction and explain why such a reduction adds additional protection from dephasing in the next section.

At this point, readers with a condensed matter or statistical physics background could ponder if there exists some order-by-disorder mechanism via which quantum fluctuations could cause this symmetry reduction. However, due to the nature of the $U(1)^\beta$ symmetry one can show that such an order-by-disorder mechanism is not possible. We refer the interested readers to appendix \ref{App_sec_beta_translations} for a more detailed discussion on order-by-disorder in these settings.


\subsection{Model $\mathbb{Z}^\beta_2$ - Dual protection model}
To obtain a discrete symmetry related to the entanglement phase, $\mathbb{Z}^\beta_2$, we need to introduce a term in the Hamiltonian which is sensitive to $\beta$. To choose a suitable term, we start by considering the classical energy function $H(\bm{\Psi})$ defined by Eq.~(\ref{def_classical_energy_functional}).
The motivation for adopting this starting point is that, as shown below in 
subsection~\ref{subsec_impossibility}, we cannot achieve the desired $\mathbb{Z}^\beta_2$ symmetry by working in the
single block case $d=1$. We shall therefore build on the physical
intuition provided by the existence of a classical limit at large values of $d$, for a system of $d$-pairs of spin and pseudospin-1/2 degrees of freedom (see sec. \ref{sec_genprinc}).

Now notice that the expectation values of any single spin or pseudospin operator does not have any sensitivity to $\beta$. Hence, we introduce a term which couples the two. We introduce the following coupling (one can consider other variations of $(\bm{S} \cdot \bm{P})^2$ terms as well)
\begin{equation}
    \bar{H}(\bm{\Psi}) = H_C(\bm{\Psi}) + v \, \langle \hat{S}_x\hat{P}_x \rangle_{\bm{\Psi}}^2
    \label{mod1a}
\end{equation}
Note that importantly, the above is the \emph{classical} energy function, where all the terms are to be interpreted as symbols (functions) of equivalent quantum operators under the geometric quantization scheme (See appendix \ref{subsec_geomquant} for details and eq. \ref{optoexp} for notation). More explicitly, as mentioned in sec. II A, quantization of the above is achieved via generalized coherent states and Schwinger bosons. So the above energy function can be thought of as the classical large-$d$ limit of a quantum mechanical Hamiltonian.

Such a coupling term is naturally sensitive to the phase $\beta$. To see this explicitly, we write

\begin{equation}
\begin{split}
    \langle \hat{S}_x\hat{P}_x \rangle_{\bm{\Psi}} = \underbrace{\rm{cos}^2\frac{\alpha}{2}\,\langle \phi | (\sigma_x \otimes \mathbb{1})(\mathbb{1} \otimes \sigma_x)| \phi \rangle}_{\text{(i)}} \\ \mbox{} + \underbrace{\rm{sin}^2\frac{\alpha}{2}\,\langle \chi | (\sigma_x \otimes \mathbb{1})(\mathbb{1} \otimes \sigma_x)| \chi \rangle}_{\text{(ii)}} \\ \mbox{}
    + \frac{e^{-i\beta}}{2} \underbrace{\rm{sin} \alpha \,\langle \chi | (\sigma_x \otimes \mathbb{1})(\mathbb{1} \otimes \sigma_x)| \phi \rangle}_{\text{(iii)}} \\ \mbox{}
    + \frac{e^{i\beta}}{2} \underbrace{\rm{sin} \alpha \,\langle \phi | (\sigma_x \otimes \mathbb{1})(\mathbb{1} \otimes \sigma)| \chi \rangle}_{\text{(iv)}}
    \label{spx}
\end{split}
\end{equation}
where we have used $\ket{\Psi} = \rm{cos}\frac{\alpha}{2}\ket{\phi} + e^{i \beta}\rm{sin}\frac{\alpha}{2}\ket{\chi}$ with $\ket{\phi} = \rm{cos}(\alpha/2) \ket{\phi_S}\otimes \ket{\phi_P}$ and $\ket{\chi} = \rm{sin}(\alpha/2) \ket{\chi_S}\otimes \ket{\chi_P}$. After some algebra and using the expressions for $\ket{\varphi_{S(P)}},\ket{\chi_{S(P)}}$ given in eq. (\ref{schmidtdec}), we get
\begin{equation}
\begin{aligned}
    &\langle \hat{S}_x\hat{P}_x \rangle_{\bm{\Psi}} = \rm{sin}\theta_S \,\rm{sin} \theta_P \,cos \varphi_S \, cos \varphi_P + \mbox{} \\ &\rm{sin} \alpha \bigg[ \rm{sin}^2\frac{\theta_S}{2}\,\rm{sin}^2 \frac{\theta_P}{2}\,\rm{cos}(2\varphi_S + 2\varphi_P - \beta) + \mbox{} \\ &\rm{cos}^2 \frac{\theta_S}{2}\,\rm{cos}^2\frac{\theta_P}{2}\,\rm{cos}(\beta)
    -\rm{sin}^2 \frac{\theta_S}{2}\,\rm{cos}^2 \frac{\theta_P}{2}\,\rm{cos}(2\varphi_S - \beta)- \mbox{} \\
    &\rm{cos}^2 \frac{\theta_S}{2}\,\rm{sin}^2 \frac{\theta_P}{2}\,\rm{cos}(2\varphi_P - \beta)\bigg]
    \label{sxpx}
\end{aligned}
\end{equation}
The first line above comes from $(i)+(ii)$ and is hence independent of $\beta$. The $\beta$-dependent terms come from $e^{-i\beta}(iii)+e^{i\beta}(iv)$. However, as we see from the above expression, an $\langle \hat{S}_x\hat{P}_x \rangle_{\bm{\Psi}}$ term is not sufficient to obtain a $\mathbb{Z}^\beta_2$ symmetry as the above expression will not be invariant under $\beta \rightarrow \beta + \pi$, except perhaps at a set of measure zero points.

However, the coupling in Eq.~(\ref{mod1a}) is the square of the right hand side in Eq.~(\ref{spx}). On expanding one gets terms independent of $\beta$, terms proportional to $e^{\pm i\beta}$ and those proportional to $e^{\pm 2i\beta}$. To get our required $\mathbb{Z}^\beta_2$ symmetry, we need the terms proportional to $e^{\pm i\beta}$ to vanish and those proportional to $e^{\pm 2i\beta}$ to be non-zero. Elementary algebra shows that this is equivalent to requiring that $(i)+(ii) = 0$ and $(iii),(iv) \neq 0$ (see Eq.~(\ref{spx})). Such a condition can be achieved if either $\theta_P = 0$ or $\theta_S = 0$ for generic values of all other angles. Physically, this imposes the requirement that the ground-state configuration has no in-plane pseudospin/spin component. 
This condition holds for the minima of the
$H_C(\bm{\Psi})$ piece, because of the Zeeman term, as we have seen
in detail in subsection~\ref{subsec_class_desc}. Adjusting $\beta$
such that $\langle \hat{S}_x\hat{P}_x \rangle_{\bm{\Psi}}$ vanishes
allows to minimize simultaneously both terms in the energy function
given in Eq.~(\ref{mod1a}) provided that $v>0$. Its ground-state is
doubly degenerate, and it exhibits the desired $\mathbb{Z}^\beta_2$ symmetry.
Importantly, note that this effective symmetry emerges only in a low energy sector,
in which the $(i)+(ii)$ contribution remains negligible, compared
to the $e^{-i\beta}(iii)+e^{i\beta}(iv)$ term.

What is the corresponding operator (to leading order in $1/d$) in the Schwinger boson Hilbert space for the symbol  $\langle \hat{S}_x\hat{P}_x \rangle_{\bm{\Psi}}^2$ in $\mathbb{C}$P(3)?  After defining $\psi_j$ components so that
$\psi_1$ corresponds to the $\uparrow_S \uparrow_P$ configuration, 
$\psi_2$ to $\uparrow_S \downarrow_P$,
$\psi_3$ to $\downarrow_S \uparrow_P$ and 
$\psi_4$ to $\downarrow_S \downarrow_P$ and setting $\bm{\Psi}^{\dagger}\bm{\Psi}=1$, one can write:
\begin{equation}
   \langle \hat{S}_x\hat{P}_x \rangle_{\bm{\Psi}}  = \psi_1^* \psi_4 + \psi_2^* \psi_3 + \psi_3^* \psi_2 + \psi_4^* \psi_1
\end{equation}
Hence we are searching for the operator corresponding to the covariant symbol $(\psi_1^* \psi_4 + \psi_2^* \psi_3 + \psi_3^* \psi_2 + \psi_4^* \psi_1)^2$. Using the dictionary of geometric quantization on $\mathbb{C}$P(3) projective space (see appendix \ref{subsec_geomquant} for procedure and primer on geometric quantization), we can show that the leading order operator is $(a_{1}a^{\dagger}_4 + a_{2}a^{\dagger}_3 + \rm{h.c})^2/d^2$. Therefore the quantum Hamiltonian with the required $\mathbb{Z}^\beta_2$ symmetry is given by
\begin{equation}
\hat{H}_{Q1} = \hat{H}_{Q0} + (a_{1}a^{\dagger}_4 + a_{2}a^{\dagger}_3 + \rm{h.c})^2/d^2
\end{equation}
where $\hat{H}_{Q0}$ is that of Eq.~(\ref{hqham}).

\subsubsection{Impossibility for d=1 $\mathbb{Z}^\beta_2$ construction}
\label{subsec_impossibility}
We have shown how to construct a $\mathbb{Z}^\beta_2$ entanglemon from a one-site model by going to higher dimensional representations of SU(4), or in other language going away from $d = 1$. The minimal construction for such a model requires $d = 2$. We explain this obstruction for such a $d=1$ construction in this section. Readers familiar with superconducting circuits can think of an analogous roadblock (although for different underlying reasons) in protection from both, depolarization and dephasing, in single-mode circuits.

 For $d = 1$ we have the fundamental representation of SU(4) and the Schwinger boson Hilbert space has dimension $\mathrm{dim}(\mathcal{H}_{SB}) = 4$. In simpler language and referring to our general principle of construction in sec. \ref{sec_genprinc}, for $d=1$, we only have a single spin and pseudospin-1/2 degree of freedom.
 In this representation, the expectation value of any operator (or Hamiltonian) $\hat{O}$, no matter how complicated, has the same structure as the right hand side of Eq.~(\ref{spx}) so it
 cannot contain any $e^{\pm i 2\beta}$ term. This implies that the fundamental representation of SU(4) cannot host the required $\mathbb{Z}^\beta_2$ symmetry which is not part of a larger symmetry group ( $U(1)^{\beta}$ for this specific case). One needs to go to higher-dimensional representations, i.e, add atleast one more pair of spin and pseudospin-1/2s, to tame the quantum fluctuations along the $\beta$ direction and to carve out such a double well-like potential for the ground state associated with the model $\mathbb{Z}^\beta_2$ entanglemon. 

\subsubsection{Dual protection from  noise}
\label{subsec_full_protection}
The $Z_{2\beta}$  entanglemon model is modified from an anharmonic oscillator for the $U(1)_\beta$ case to a double-well like structure corresponding to the two minima for the $\mathbb{Z}^\beta_{2}$ symmetry. Encoding information in these two states induced by the $\mathbb{Z}_{2}^{\beta}$ symmetry, as we show below, enhances protection from any depolarization \textit{and} also enables protection from dephasing errors. As a reminder, this heuristic analysis relies on symmetric noise which acts of all $d$ pairs similarly. See sec VI for a more detailed analysis of other kinds of possible noise.

Let us first provide a heuristic argument for this enhanced protection.
This heuristic has become relatively
standard, see e.g.~\cite{gottesman2001encoding,douccot2012physical,beytransm}.
To protect against bit-flips, we need the wave-functions associated to the 
two qubit states $\ket{0} =(\ket{\beta_0} + \ket{\beta_\pi})/\sqrt{2}$ and $\ket{1} =(\ket{\beta_0} - \ket{\beta_\pi})/\sqrt{2}$ to have disjoint supports in some representation. Here we are referring to
the notations introduced in Fig.~\ref{figmodels}.
The low energy states $\ket{\beta_0}$ and $\ket{\beta_{\pi}}$ correspond to opposite (with respect to the centre of the $\alpha$ orbit) points.

If we neglect quantum tunneling effects, these two states are
degenerate, and can be used as a natural basis for the qubit logical Hilbert space. For finite values of $d$, this degeneracy is lifted by tunneling,
and the even ($\ket{0}$) and odd ($\ket{1}$) combinations  are selected. If $G$ is the
variable conjugate to $\beta$, the $\ket{0}$ state is invariant under the
$\beta$ translation by $\pi$, so it is contained in the subspace in which
eigenvalues of $\hat{G}$ are \emph{even integers}. By contrast, the
$\ket{1}$ state changes into its opposite under the
$\beta$ translation by $\pi$. Therefore, it is contained in the subspace in which
eigenvalues of $\hat{G}$ are \emph{odd integers}. As we have argued in~\ref{sub_noise_protection_dep}, single spin or pseudospin operators 
do not induce transitions between eigenvectors of $\hat{G}$
with distinct eigenvalues. As a result, their matrix elements between
the $\ket{1}$ and the $\ket{0}$ states vanish.

The strong localization of the two components
$\ket{\beta_0}$ and $\ket{\beta_\pi}$ near two maximally distinct values of the $\beta$
phase, emphasized in Fig.~\ref{figmodels}, gives these states a large spread in the conjugate quantum variable $\hat{G}$. Because of this spread, the qubit states $\ket{0}$ and  $\ket{1}$ are now indistinguishable by measuring $\hat{G}$. The diagonal matrix elements of $\hat{S}_{x,y,z}$ and $\hat{P}_{x,y,z}$ operators are the same for both the
$\ket{0}$ and $\ket{1}$ states.

Therefore, our simple model successfully realizes a qubit protected from \emph{both} depolarization and dephasing. We would like to emphasize once again the generality of this construction. The models we present are completely platform agnostic, and only require the presence of a coupled array of $\mathbb{C}$P(3) units. In the next section we describe how this can be realized in four quantum computing hardware platforms. Such a list is  by no means exhaustive, we hope interested readers with expertise in more specific aspects of quantum hardware can expand this list. Further, we go beyond the 
assumption that the noise involves only uniform modes
of the environment, and present a more detailed  characterization of the effects of noise on the model $\mathbb{Z}^\beta_2$ - entanglemon.

\section{Implementations of $\mathbb{Z}^\beta_2$ entanglemon --- General principle}
\label{subsec_impl_Mod_1} 
Now that we've achieved a complete description of the symmetric subspace and the noise protection within that subspace for the two entanglemon models, we shall describe how such subspaces can be embedded in quantum hardware. The generality of our construction also allows for a wide variety of approaches towards such hardware implementations. Some of these approaches can viewed through the lens of error correction and implementing ideas from repetition codes and stabilizer codes to achieve the ground states required for entanglemons \cite{gottesman1997stabilizer,CScode,steane1996multiple}. 
Before going to hardware specific content, we  introduce a general principle for the construction of the $\mathbb{Z}^\beta_2$ entanglemon a special case of which has connections with mentioned error correction schemes. In the next section we shall use these connections to present two concrete implementations on superconducting circuits and trapped ions. In the next section we also sketch the possibilities for solid state platforms, some of which could go beyond hardware implementations of repetition and stabilizer-like error correction schemes.

Let us assume that we have a collection of $d$ identical sites, and that each 
of them hosts a pair of  two level systems. We also assume that the local site Hamiltonian has a nearly doubly degenerate ground-state, well separated in energy from the two remaining excited states, such that this ground-state
subspace can be generated by a pair of states of the form $\ket{AA'}_i$ and
$\ket{BB'}_i$, where $\ket{A}_i$, $\ket{B}_i$ (resp. $\ket{A'}_i$, $\ket{B'}_i$)
is an arbitrary orthonormal basis for the first (resp. the second)
two level subsystem at a given site $i$.

Note that having a basis of this form is a rather strong constraint for a two-dimensional subspace included in the Hilbert space for a pair of spins 1/2.
The latter space being four-dimensional, the family of its two dimensional subspaces forms the Grassmannian manifold $\rm{Gr}(2,4)$,
whose complex dimension is 4, so it can be described (locally) by 8 real parameters. But the above collection of two-dimensional
sub-spaces is described only by 4 real parameters, that arise
from choosing one point on the spin Bloch sphere for the $(\ket{A},\ket{B})$ pair of orthonormal states, and one point on the pseudospin Bloch sphere for the $(\ket{A'},\ket{B'})$ pair. This constraint ensures that the two-fold degenerate single site ground-state can host a pair of partially entangled states, each of the form 
$\rm{cos}\frac{\alpha}{2} \ket{AA'}+e^{i\beta}
\rm{sin}\frac{\alpha}{2}\ket{BB'}$, and differing only by the
value of the entanglement phase $\beta$.

We wish to stabilize fully symmetric states of the form
$\otimes_{i} \ket{\alpha,\beta}_i$, where 
$\ket{\alpha,\beta}_i = \rm{cos}\frac{\alpha}{2} \ket{AA'}_i+e^{i\beta}
\rm{sin}\frac{\alpha}{2}\ket{BB'}_i$ and
the entanglement phase $\beta$ can be either 0 or $\pi$ as in Fig. \ref{figmodels}.
On the entanglement Bloch sphere, these two states can be viewed
as coherent states pointing along directions whose spherical 
coordinates are $(\alpha,0)$ and $(\alpha,\pi)$. Let us introduce
an entanglement unit vector $\bm{n}_E$ in three dimensional
Euclidean space. The two coherent states minimize an energy function
$E(\bm{n}_E)=-\lambda n_E^z - \gamma (n_E^x)^2$, provided
$\lambda$ and $\gamma$ are positive and satisfy $\frac{\lambda}{\gamma}=\cos{\alpha}$. 

To understand this in the Schwinger boson language, as in the two models in the above sections, let us choose a Schwinger boson basis
such that $a_1^{\dagger}$ creates state $\ket{AA'}$,
$a_2^{\dagger}$ creates state $\ket{BB'}$, 
$a_3^{\dagger}$ creates state $\ket{AB'}$, and
$a_4^{\dagger}$ creates state $\ket{BA'}$.
The above energy functional, $E(\bm{n}_E)$ corresponds to the quantum Hamiltonian (to leading order
in $1/d$):
\begin{equation}
    \hat{E}_{Qd}= - \frac{\lambda}{d}(a_1^{\dagger}a_1 - a_2^{\dagger}a_2) - \frac{\gamma}{d^2}(a_1^{\dagger}a_2 +a_2^{\dagger}a_1)^2
\end{equation}
The first term can be added to the single site Hamiltonian, because
it is quadratic in elementary bosonic operators. The second term is
quartic, so it corresponds to an all to all pairwise interaction between
sites. In the extended Hilbert space, of total dimension $4^d$, this
interaction term may be written as (in first quantized language)
\begin{equation}
    \hat{E}_{\rm{int}}= -\gamma\sum_{1\leq i,j \leq d} \hat{U}_{ij}
\end{equation}
where the two site operator $\hat{U}_{ij}$ is defined by
$\hat{U}_{ij}\ket{11}_{ij}=\ket{22}_{ij}$, 
$\hat{U}_{ij}\ket{22}_{ij}=\ket{11}_{ij}$,
$\hat{U}_{ij}\ket{12}_{ij}=\rho \ket{12}_{ij}+\nu \ket{21}_{ij}$,
and $\hat{U}_{ij}\ket{21}_{ij}=\nu \ket{12}_{ij}+ \rho\ket{21}_{ij}$,
with the conditions $\rho+\nu =1$ and $\nu$ positive, in order to
give an energy penalty to the anti-symmetric state 
$\ket{12}_{ij}-\ket{21}_{ij}$.

Because we have only a single quartic term, we shall describe a simpler special case obtained when the two single site states 
$\ket{\alpha,\beta=0}$ and $\ket{\alpha,\beta=\pi}$ are orthonormal,
which requires $\alpha=\pi/2$, so that the corresponding states are
maximally entangled. Also, $\alpha = \pi/2$ corresponds
to setting $\lambda=0$ in $E(\bm{n}_E)$. Now,
let $\tau_i$ be the operator
that exchanges states $\ket{AA'}_i$ and $\ket{BB'}_i$. The two eigenvectors
of this operator have the form $\ket{AA'}_i + e^{i\beta}\ket{BB'}_i$ with
eigenvalue 1(-1) for $\beta =0(\pi)$. 
In order to stabilize fully symmetric states of the form
$\otimes_{i}(\ket{AA'}+e^{i\beta}\ket{BB'})_i$,
in which each site is in a maximally entangled superposition of 
states $\ket{AA'}$ and $\ket{BB'}$, and the entanglement phase $\beta$ is either 0 or $\pi$, it is natural to consider an anisotropic generalized ferromagnetic coupling with an Ising-type $\mathbb{Z}_2$ symmetry:
\begin{equation}
H=-\sum_{1\leq i,j \leq d, i \neq j} \tau_i \, \tau_j   
\end{equation}
This is the setting that will be used in sections~\ref{subsec_ions}, \ref{subsec_circuits} below.

\section{Hardware Implementations of Entanglemon Qubits}
\label{sec_implementations}

Having described the simple models in  detail and general principle for their construction, we now turn to implementations of entanglemons on current hardware platforms for quantum computers. The diverse nature of these platforms highlights the generalization of the entanglemon construction. We focus on four platforms. Two of these are synthetic, ion traps and superconducting circuits, and two are solid-state, namely quantum dots in graphene/silicon and quantum Hall skyrmions in graphene. Our proposals are just starting points for realizations of protected qubits in these platforms, we hope the generality of our construction will inspire further platform specific, more detailed and hopefully smarter constructions of such protected qubits.

\subsection{Trapped ions}
\label{subsec_ions}
\begin{figure}
    \centering
    \includegraphics[scale = 0.3]{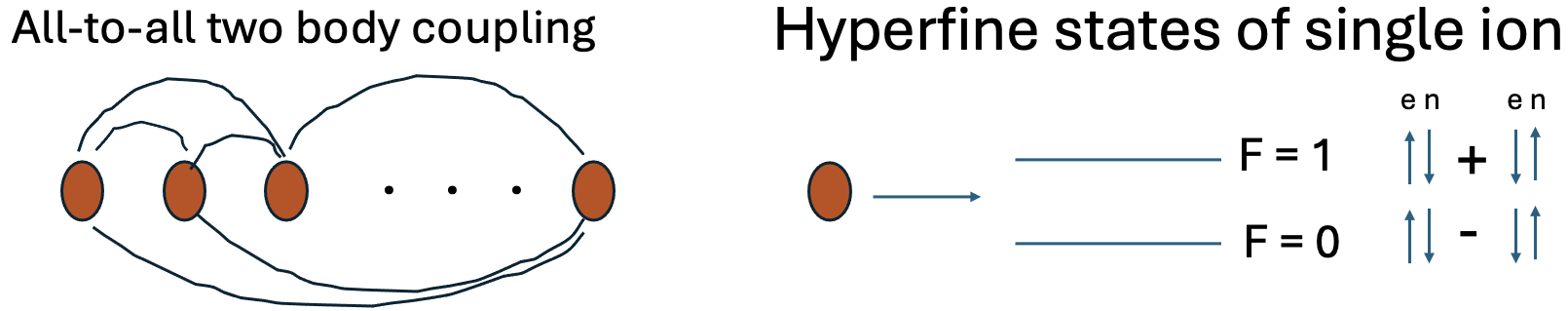}
    \caption{Constructing entanglemon qubits from the hyperfine states of trapped ions such as Yb. The hyperfine states correspond to two entangled states between electronic and nuclear spin. The coupled array of such $\mathbb{C}$P(3) units is then realized by coupling to the motion modes of the ion array, which induces a "spin"-dependent (where spin up and down are the two hyperfine states) pairwise and all-to-all coupling resulting in a collective state of all ions in the $F = 0,m_F = 0$ state or all in the $F = 0,m_F = 1$ state. The minimal hardware requirement to realize the noise-immunity features of model $\mathbb{Z}^\beta_2$ is two ions, i.e $d = 2$. More ions can be coupled to increase noise protection. }
    \label{fig:enter-label}
\end{figure}

Trapped ions are one of the oldest and most promising platforms for quantum computing. Unique to such trapped ion platforms are the relative robustness of ion-based qubits as compared to other hardware platforms and the possibilities to generate long-range interactions resulting in designer many-body Hamiltonians \cite{Ciracqub,bruzewicz2019trapped,Monrev}.

For realizing the basic unit in trapped ions we propose hyperfine qubit states in which the lowest energy levels are that of the hyperfine states with electronic and nuclear spin entangled. As an example consider the Ytterbium ion $\leftindex^{171}{\text{Yb}}^+$ (or Cadmium), a common ion used in several ion-trap implementations due to nuclear spin-1/2  \cite{monytb,blinov2004quantum}. The two "clock" hyperfine states used for most such implementations are the $F = 1, m_F = 0$ and $F= 0, m_F = 0$. These two states are exactly the maximally entangled states of electronic and nuclear spins, triplet and singlet respectively. Hence, the first requirement of on-site spin pseudospin entanglement for the entanglemon construction is naturally present in such platforms. We consider a chain of such ions with non-zero electronic and nuclear spin, each of which are pairwise coupled via some two-body interaction term. 

We can use the general discussion of the previous subsection, 
where the local state  $\ket{AA'}_i + \ket{BB'}_i$ (with $\tau_i = 1$)
corresponds to $\ket{F = 1,m_F = 0}_i$, 
and the state $\ket{AA'}_i - \ket{BB'}_i$ (with $\tau_i = -1$) corresponds to 
$\ket{F = 0,m_F = 0}_i$.
To obtain the equivalent of the large-$d$ limit, we are led then to propose a realization of such two body interactions using standard coupling of ions with motional degrees of freedom of the ion array which generate an Ising-like interaction term between $\tau_i$ "spins". 
Such a "spin"-dependent coupling is a standard approach in the quantum simulation of trapped ions \cite{Monrev}.

Once we have a coupled array of $d$ such trapped ions in which all the ions are in either the $ \ket{F = 0,m_F = 0}$ or all are in the $ \ket{F = 1,m_F = 0}$ (ferromagnetic state), we have the required permutation symmetry required for the entanglemon construction and the two collective states for the entanglemon doublet. 

\subsection{Superconducting circuits}
\label{subsec_circuits}
\begin{figure}
    \centering
    \includegraphics[scale = 0.27]{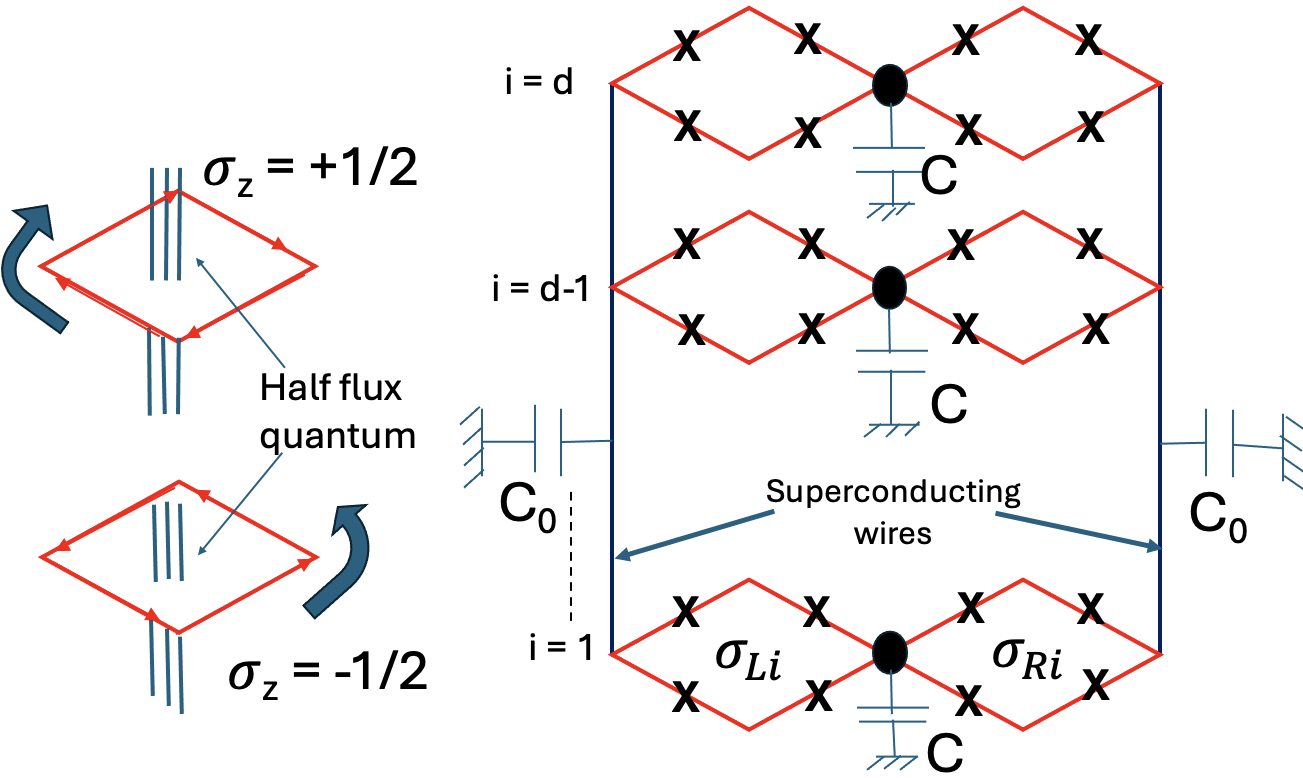}
    \caption{A superconducting circuit implementation of the entanglemon. The basic unit is a set of coupled Josephson junctions and each such unit is connected by a superconducting wire. The chirality of the supercurrent flow across each rhombus acts as an effective spin-1/2, hence the two rhombi set acts as an effective $\mathbb{C}$P(3) unit. The minimal hardware requirement for noise-protection is $d = 2$, i.e, only two rungs in the ladder. Immunity to noise can be further increased on increasing $d$.}
    \label{fig:supent}
\end{figure}
Another synthetic platform with great promise for quantum computing, especially for digital quantum computers, is presented by superconducting circuits. These circuits utilize the phase difference across Josephson junctions as the physical degree of freedom for the qubit. Starting from the Cooper pair box there has been tremendous progress in designer qubits with better error protection in this area leading to modern qubit designs such as the transmon and the fluxonium and many others \cite{martindev,koch2007charge,manufluxonium,brooks0pi,devoret2013superconducting,you2005superconducting,schoelkopf2008wiring,beytransm}. 

Here, we propose an implementation of entanglemon qubits using Josephson junction arrays similar in spirit to earlier approaches to realize protected qubits \cite{douccot2012physical}. The approach for this implementation follows a similar route as that for the ion traps. We consider a ladder-like geometry of an array of two coupled junctions, as shown in fig. \ref{fig:supent}. Each rhombus in fig. \ref{fig:supent} is threaded by 1/2 flux quantum, resulting in a two-fold degenerate energy landscape and an effective two level system. The eigenvalue of $\sigma^z$ then corresponds to the chirality of the supercurrent flow pattern around the rhombus. 

Each rung of the ladder has two rhombi and the rungs are connected to two large islands (vertical wires in fig. \ref{fig:supent}) on the left and right side of the ladder. Their capacitance to the ground is expected to be much larger than the one of the intermediate island connecting two rhombi, i. e $C_0 \gg C_1$.

Here, each horizontal rung plays the role of a single site
in the general discussion in subsection~\ref{subsec_impl_Mod_1}. The two subsystems are the left and the right rhombus on a given rung.
In such a construction we get the following low-energy Hamiltonian
\begin{equation}
    H_{eff} = -E_{\mathrm{flip}} \sum_{1 \leq i \leq N} \sigma_{Li}^x \sigma_{Ri}^x - E_{\mathrm{phase}} \sum_{1 \leq i \neq j \leq N} \tau_i \tau_j
\end{equation}
where the first term corresponds to $\pm \pi$ phase changes in the connecting island of each rung due to tunneling events whereas the second term occurs since the phase difference across a pair of rhombi is determined by $\sigma_{Li}^z \sigma_{Ri}^z = \tau_i$. The vertical superconducting wires then ensure that the phase differences across all rungs coincide.

The connection with the general discussion in subsection~\ref{subsec_impl_Mod_1} is ensured by the fact that the
single site Hamiltonian, namely $-\sigma_{Li}^x \sigma_{Ri}^x$ has 
two degenerate ground states $\ket{++}_i$ and $\ket{--}_i$,
where $\sigma_{ai}^x \ket{\pm}_{ai} = \pm \ket{\pm}_{ai}$ and
$a=L$ or $a=R$. It is then clear that $\tau_i$ exchanges $\ket{++}_i$ and $\ket{--}_i$. The Hamiltonian has all mutually commuting terms and hence represents a stabilizer code formulation but directly implemented in the construction of a qubit and not for error correction. 

\subsection{Quantum dots in graphene and silicon}
Let us now move away from synthetic platforms and focus on material (solid-state) platforms for quantum computing. In this section we focus on quantum dot platforms in materials like Silicon or graphene. Quantum dots in these materials as a platform for quantum computing are very appealing from the scalability point of view due to the vast amount of knowledge and expertise already present for materials like silicon or graphene. Pioneering work in the early late 90s, early 2000s proposed several spin to qubit architectures, such as the Loss-DiVincenzo proposal \cite{LDqubit} for a qubit which utilizes  single electronic spin states and Kane's proposal \cite{kane1998silicon} for a nuclear spin based qubit and many more (see \cite{spinqubrev} for a review of current state of the art). 

Our discussion in this section and next is at a more speculative level than the previous implementation proposals. We hope that people with more expertise on these platforms can provide more concrete implementations along the lines suggested in this work. We first describe possibilities of realizing the basic unit of model $U(1)^{\beta}$, and then subsequently describe ways of coupling these units in order to realize the large $d-$limit of the Schwinger boson approach.

 We consider an electron in a single quantum dot, which has a spin and a valley degree of freedom. Such is the case for monolayer graphene (or bilayer where valley is replaced by layer) and also for Silicon which has three equivalent pseudospin degrees of freedom which can be reduced to a single valley degree of freedom in the presence of uniaxial strain. 

The requirements for model $U(1)^{\beta}$ are minimal, one needs a $S^z$ term which for both these cases can be induced by an out-of-plane magnetic field. In addition, one needs pseudospin anisotropy terms. These pseudospin terms could arise as a consquence of short-range interactions as in the lowest Landau level of graphene \cite{kharithall} (whose anisotropy functional has the same form as model $U(1)^{\beta}$) or in the presence of other kinds of strain for Silicon or other quantum dot platforms. Usually in these platforms, charge noise is ubiquitous and leads to depolarization errors, hence the model $U(1)^{\beta}$ construction protects from all such errors and other depolarization mechanisms as well.

To get the large-$d$ limit, we consider $d$ such quantum dots, coupled with each other with a SU(4)-invariant nearest-neighbour exchange interaction $J|\langle \Psi^i | \Psi^j \rangle |^2$ with $J<0$ (Say in a triangle for $d = 3$),  where $\ket{\Psi^i}$ is the SU(4) spinor on site $i$. Such a construction implies that classically all-sites have the same low-energy doublet as model $U(1)^{\beta}$ and the low-energy quantum states are the coherent states given by the Schwinger bosons. In such a construction $d$ becomes the number of such quantum dots. 

At present, we are not able to think of a direct realization of model $\mathbb{Z}^\beta_2$ which has protection from depolarization and dephasing based errors. We hope the theoretical simplicity of model $\mathbb{Z}^\beta_2$ will motivate people working in the quantum dot community to come up with better implementations for such protected entanglemons. One could think of using singlet-triplet qubits \cite{levyst} with one electron per qubit in a double quantum dot structure as the basic unit and then ferromagnetically coupling (in the generalized sense, i.e singlet $\rightarrow \ket{\tau=-1}$, triplet $\rightarrow \ket{\tau=1}$) each set of double quantum dots in the nearest neighbour way as in the superconducting circuits section. Our lack of expertise in this area limits us to say anything more detailed about such coupling. There have been proposals of two qubit couplings using capacitive coupling which could probably be useful in this context \cite{taylorcap,nichol2017high}. A detailed implementation of this sort is an interesting avenue for future work and would parallel the approach taken for the ion-traps and superconducting circuits.

We did consider a completely different model of a  double quantum dot with antiferromagnetic exchange interactions between the dots, as a route to achieve $\mathbb{Z}^\beta_2$ without $U(1)^{\beta}$ symmetry. However, we ran into a roadblock: even if we could break the $U(1)^\beta$ symmetry associated with $(\beta^a,\beta^b) \rightarrow (\beta^a+\theta,\beta^b+\theta)$ ($a,b$ are the site-labels) down to a $\mathbb{Z}^\beta_2$ subgroup with $(\beta^a,\beta^b) \rightarrow (\beta^a+\pi,\beta^b+\pi)$ there was still a remnant antisymmetric U(1) symmetry, i.e invariance under $(\beta^a,\beta^b) \rightarrow (\beta^a+\theta,\beta^b-\theta)$,
for arbitrary $\theta$. Hence the discrete $\mathbb{Z}^\beta_2$ symmetry is still accompanied by a larger $U(1)^{\beta}$ symmetry in this model. We include details of this model in the appendix as exchange coupling models are widely implemented in quantum dot experiments, and this simple model could serve as a base for further improvements and perhaps a true discrete $\mathbb{Z}^\beta_2$ symmetry without any additional U(1) group. Moreover, such a model and its modifications would also be interesting to study the physical properties of $\mathbb{C}$P(3) antiferromagnets.

\subsection{Quantum Hall skyrmions in graphene}
An immediate extension of the simple models where we employ $d$ identical copies to access the Schwinger boson Hilbert space, is to consider an extended object where every site is only slightly different from its neighbours. This arises naturally in solid-state platforms in the form of skyrmion textures. Skyrmions in chiral magnets as qubits were proposed in inspiring earlier work \cite{skyrmqub}. To have any kind of intrinsic protection from error as in the entanglemon construction, we would require a topological texture of $\mathbb{C}$P(3) degrees of freedom. Such a texture could occur as a metastable configuration in the quantum Hall ferromagnetic regime in graphene and have been studied in detail in recent works \cite{Doucotent,LianPRL}.

Consider monolayer graphene in the zeroth Landau level. At $\nu = 1$ filling, the ground state is a quantum Hall ferromagnet where the zeroth Landau level is fourfold degenerate (each labelled by spin and valley) and Coulomb interactions are approximately SU(4) symmetric \cite{goerbig2006electron,aliceaprb}. The standard anisotropies relevant for monolayer graphene have exactly the form of $H_C$ \cite{kharithall}. Excitations out of this quantum Hall ferromagnet are skyrmions \cite{sondhiskyr}. In the $u_p > u_z > \Delta/2$ regime, one gets skyrmions in which the spin and valley degrees of freedom are entangled and the entanglement Bloch vector covers the full entanglement sphere - these have been termed entanglement skyrmions \cite{Doucotent,LianPRL}.

Are there conditions under which such entanglement skyrmions can host a discrete $\mathbb{Z}^\beta_2$ symmetry? To examine this we note that the criterion for such a global $\mathbb{Z}^\beta_2$ invariance of the exchange energy (in the continuum limit) can be reduced to the condition
\begin{equation}
   \underbrace{ \int d^2 \bm{r} \big( e^{i\beta(\bm{r})} \langle \dot{\phi} (\bm{r})| \dot{\chi}(\bm{r}) \rangle}_{\text{z}} +e^{-i\beta(\bm{r})} \langle \dot{\chi} (\bm{r})| \dot{\phi}(\bm{r}) \rangle \big) = 0
\end{equation}
Here, the notation refers to the Schmidt decomposition in eq.~(\ref{schmidtdec}) of the four component spinor $\ket{\Psi(\bm{r})}$
describing the skyrmion texture at position $\bm{r}$ in the plane.
More precisely, $\ket{\phi}=\cos{\frac{\alpha}{2}} \ket{\phi_S}\ket{\phi_P}$, and $\ket{\chi}=\sin{\frac{\alpha}{2}} \ket{\chi_S}\ket{\chi_P}$. To shorten notations, we set $\dot{f}=\bm{\nabla}f$
for any scalar or spinor-valued function $f(\bm{r})$.

Let us denote the first term of the above equation by $z$, so we can write the above equation as $z+z^* =0$. 
Consider a global transformation that sends $\beta(\bm{r}) \rightarrow \beta(\bm{r})+\theta$. Under this transformation $z+z^* \rightarrow e^{i\theta}z + e^{-i\theta}z = 2 \cos{\theta}\, \rm{Re}(z) - 2 \sin{\theta}\, \rm{Im}(z)$. 
$\mathbb{Z}^\beta_2$ symmetry requires that the above be equal for $\theta = 0$ and $\pi$ which requires that Re$(z) = 0$. However, if also Im$(z) = 0$, then full $U(1)^{\beta}$ invariance is present which gives us entanglemons completely immune to depolarization but susceptible to dephasing, as we have seen in the sections above. Hence, to reduce the global symmetry from $U(1)^{\beta}$ to $\mathbb{Z}^\beta_2$, the skyrmions need to satisfy the condition Re$(z) = 0$ and Im$(z) \neq 0$.

After some algebra, using the expressions for $\ket{\phi}$ and $\ket{\chi}$ in eq. (\ref{schmidtdec}) one can show that the above condition for skyrmions can be concretely written as 
\begin{equation}
\begin{aligned}
   \text{Re}(z) &=  \int d^2 \bm{r} (\sin{\theta_S}\sin{\theta_P}\dot{\varphi_p}\dot{\varphi_S} - \dot{\theta_P}\dot{\theta_S})\sin{\alpha}\cos{\beta'} \\
&+(\sin{\theta_S}\dot{\varphi_S}\dot{\theta_P} + \sin{\theta_P}\dot{\varphi_P}\dot{\theta_S})\sin{\alpha}\sin{\beta'} = 0\\
  \text{Im}(z) &= \int d^2 \bm{r} (\sin{\theta_S}\sin{\theta_P}\dot{\varphi_p}\dot{\varphi_S} - \dot{\theta_P}\dot{\theta_S})\sin{\alpha}\sin{\beta'} \\
&-(\sin{\theta_S}\dot{\varphi_S}\dot{\theta_P} + \sin{\theta_P}\dot{\varphi_P}\dot{\theta_S})\sin{\alpha}\cos{\beta'} \neq 0\\
\end{aligned}
\label{condskyrm}
\end{equation}
where $\beta' = \beta - \varphi_P - \varphi_S$ and the $\bm{r}$ dependence of all angles is implicit and is hidden to avoid clutter. 

We know that
as $r \rightarrow \infty$, the spins and pseudospins of the entanglement skyrmion point along the z-axis, i.e $\theta_{S,P},\varphi_{S,P},\alpha = 0$ \cite{LianPRL}. Let us consider a skyrmion of small radius (size) $r_s$ (skyrmion size is easily tunable in graphene by tuning the external magnetic field). By construction, all angles and hence the terms in the brackets above are equal to 0 for $r > r_s$. Hence, we need the above conditions to hold for the integral over a disc of radius $r_s$. 

There could be some parameter regime in which such fine-tuned conditions are met and the $\mathbb{Z}_2$ version is realized, however, we leave that for future detailed work.
Such a regime would realize entanglemon qubits protected from depolarizing and dephasing similar to model $\mathbb{Z}^\beta_2$. However importantly, parity (or rotational) symmetric entanglement skyrmions possess a full $U(1)^\beta$ symmetry (where both L.H.S in eq. \ref{condskyrm} = 0 ) and can realize a qubit protected from depolarizing errors as in model $U(1)^{\beta}$.

To finalize the entanglemon construction in such a platform, we propose the two lowest energy bound states of the skyrmion as the low energy doublet. The barrier to such a proposal is obtaining a well separated doublet of such bound states that are gapped (rendering the skyrmion metastable) and lie below the magnon-continuum. Such conditions are already met for the case of a spin skyrmion in chiral magnets as shown in \cite{skyrmmodes,quantskyrm} for reasonably high magnetic fields (small skyrmions) where a doublet comprising the translational mode and the elliptic deformation mode of the skyrmion form a well separated doublet suitable for the qubit basis states. We leave a detailed characterization of $\mathbb{C}$P(3) skyrmion internal modes including material specific considerations for future work.

Our entanglemon construction from the internal modes of a skyrmion provides a proposal to form protected qubits in graphene, and possibly in other graphene-based moire platforms where a variety of pseudospin-like flavours emerge 
\cite{cao2018correlated,cao2018unconventional,khalaf2021charged,khalaf2022baby,kwan2022skyrmions,chatterjee2020symmetry}. Further, the presence and control of such skyrmions could be established via recently developed magnon scattering properties in such platforms \cite{chakrabortymagnon}. 

\section{Noise protection beyond uniform approximation and minimal hardware requirements}
\label{sec_noiseanalysis}
In this section we do a detailed qualitative analysis of the protection afforded by both entanglemon models from different kinds of noise. We group the possible sources of noise into two types: Local (asymmetric) noise, which depends on the index of the copy and symmetric noise which affects all of the $d$-copies similarly, i.e it preserves the permutation symmetry of the $d$-copies.

Sources of the former kind, for superconducting circuits, include local flux noise due to defects or impurities, which usually scale with the length of the flux loops, charge noise due to fluctuations in the gate charge of the islands etc. For quantum dot platforms various sources of charge noise fall in this category as well. Symmetric noise sources, based on our implementations, could include a noisy voltage source at the points where the left and right superconducting wires are grounded (see Fig. \ref{fig:supent}), global flux noise etc.. Whereas for solid-state platforms these could include phonon induced errors as well as fluctuations of the magnetic field. The latter sources of noise (symmetric noise) can sometimes be particularly harmful, as in standard formulations of qubits such noise that has a finite correlation length can induce correlated errors thereby making quantum error correction extremely difficult.

\subsection{Noise protection analysis for $U(1)_\beta$}

\textbf{Bit flip errors from local (asymmetric) noise}: We term local noise, the generic noise terms for which $\hat{B}_i$ depends on $i$. If the system is initially prepared in the $\ket{0}$ or $\ket{1}$ state of the $U(1)_\beta$ entanglemon the action of such noise operators induce transitions away from the symmetric Schwinger boson subspace. However, as described above, the entanglemon models are implemented by considering a microscopic Hamiltonian with all to all generalized ferromagnetic interactions which stabilize the low-energy symmetric subspace.

For such implementations, and for systems with a continuous symmetry (as is for $U(1)_\beta$) one can show on rather general grounds (see appendix E) that the gap between the lowest-energy states and the non-symmetric subspace scales as $\sqrt{d}$. As a result, any such small local noise term is \textit{exponentially} ($\sim e^{-\sqrt{d}}$) suppressed. Further, there is a clear separation between the symmetric subspace states, for finite $d$, the gap to which as seen above scales as $1/d$.

\textbf{Bit flip errors from symmetric noise}: Besides such local noise suppression due to the implementation, the special feature of the entanglemon construction is that the logical states are protected from symmetric noise terms that would induce transitions within the symmetric Schwinger boson subspace. As a reminder, these are noise terms in which $\hat{A}_i$($\hat{B}_i$) does not depend on $i$. This would include all kinds of correlated noise, that act on the $d$-copies of the spin-1/2 pairs identically. As we saw in sec. \ref{sub_noise_protection_dep}, the $U(1)_\beta$ entanglemon is protected from bit-flip errors upto \textit{first order} in such symmetric noise terms.

\textbf{Phase errors}: As mentioned in sec. \ref{sub_noise_protection_dep}, the $U(1)_\beta$ entanglemon offers no protection against any kind of dephasing errors, these errors can only be protected in the $Z_{2\beta}$ entanglemon, as we describe below.

To summarize, the $U(1)_\beta$ entanglemon is well protected from bit-flip errors but susceptible to phase-flip errors. However, the strong protection from depolarization offers great advantages in reducing the overhead for quantum error correction since good performance of the qubit can be obtained by correcting for only dephasing errrors. Moreover, since the qubit also protects from symmetric noise induced bit-flip errors, the chances of correlated bit-flip errors are reduced making error correction more feasible.

\subsection{Noise protection analysis for $Z_{2\beta}$}
In sec. \ref{subsec_full_protection} we presented a heuristic argument for the protection of the $Z_{2\beta}$ entanglemon. Now that we have presented concrete implementations of the model above, we provide a more quantitative and detailed analysis. As for the $U(1)_\beta$ case we split the analysis into asymmetric (local) and symmetric noise effects.

\textbf{Bit flip errors from local (asymmetric) noise}: Much like the $U(1)_\beta$ case, the $Z_{2\beta}$ entanglemon is also protected such noise induced bit-flips due to the nature of the implementations proposed in sec. IV and V, which result in an effective quantum Ising model formulation. However, there's a very important difference with the $U(1)_\beta$ case. In such a formulation, one can show (see appendix E) that the lowest energy states of this effective Ising model are separated in energy from the other symmetric and non-symmetric states by a gap $\sim d$, hence $\textit{all}$ noise terms (assuming they're small) that take you out of the logical subspace are \textit{exponentially} suppressed in $d$.

\textbf{Bit-flip errors from symmetric noise:}
As we mention above, in contrast to the $U(1)_\beta$ case, since the gap to other states in the symmetric subspace is $\sim d$, any symmetric noise that would couple the logical states to one of these states is \textit{exponentially} suppressed in $d$. Moreover, even the bit-flip errors within the computational subspace are \textit{exponentially} suppressed since, as we saw above, the logical states are defined as states in which all $\tau_i = +1$ and $-1$ respectively for $i = 1,..,d$. Hence, one needs to go to order $d$ in perturbation theory to connect the two states.

\textbf{Phase errors due to local noise}: In addition to the enhanced protection against bit-flip errors, the $Z_{2\beta}$ entanglemon is also protected from dephasing. To see this, now consider the effect of $\sigma_{Li}^z$ and $\sigma_{Ri}^z$ operators. Since these operators commute with $\tau^z_i$, they do not induce any depolarization in the basis where all $\tau^z_i$s are equal. A
single application of a $\sigma_{Li}^z$ or a $\sigma_{Ri}^z$ operator on the ground-state doublet creates an excited state on row $i$, so we are protected to first order against such noise terms.
However, second order operators $\sigma^z_{Li} \sigma^z_{Ri}$ commute with
all $\tau_j^z$ operators and will generate dephasing in the basis
defined by eigenvalues of these operators.
So we get dephasing at second order in local noise amplitudes.

\textbf{Phase-errors from symmetric noise:}
As we discussed in sec. \ref{subsec_full_protection}, the logical states of the $Z_{2\beta}$ entanglemon are localized around two maximally distinct values of $\beta$ and hence have a broad spread in the conjugate variable $\hat{G}$. Hence, using the same reasoning as is done in the $0-\pi$ qubits \cite{brooks0pi}, these states are \textit{exponentially} (in $d$) protected from dephasing induced errors caused by symmetric noise terms.

Hence, the $Z_{2\beta}$ entanglemon realizes an extremely well protected qubit. Not only does it provide exponential protection againt both symmetric and asymmetric noise, thereby greatly reducing the chances of any bit-flip errors, it is also very well protected from phase errors. For reasons mentioned earlier, this makes the $Z_{2\beta}$ entanglemon a very good candidate for a protected qubit which has the promise to greatly reduce the overhead for error correction and also substantially increases the chances for error correction to work at scale.

\subsection{Hardware considerations}
Finally, we comment on the hardware requirements and efficiency of the entanglemon construction and compare the requirements with those of other protected qubit formulations.

The only lower bound comes from the "no-go result" of $d=1$ to realize immunity from both dephasing and depolarization. Hence, a minimal requirement for the $Z_{2\beta}$ entanglemon construction is $d=2$, i.e two ions with hyperfine states, two rows of a pair of superconducting islands or possibly two quantum dots. Of course, as we saw above, increasing $d$ will increase the protection further but already at modest $d=2-4$ the entanglemon construction should provide protection from both depolarization and dephasing.

Our analytical analysis has been largely semiclassical hence, one could expect stronger quantum effects at lower values of $d$ which could change the energetics of the entanglemon qubit levels. However, since as we saw in sec. \ref{sec_implementations}, the entanglemon levels arise from a general ferromagnet construction or possibly from $\mathbb{C}$P(3) skyrmions arising from generalized ferromagnets. In both these cases, quantum (zero-point) fluctuations don't change the qualitative picture from the semiclassical analysis \cite{douccot2018zero}.

Finally, we believe that both the entanglemon models and the implementations we present in this work represent scalable protected qubits. The requirement for all to all interactions and relatively small values of $d \sim 2-4$ are achievable in the implementations under current experimental capabilities \cite{Ioffehard}. Further, almost all other protected qubit designs in the superconducting circuit literature (where most such proposals exist) rely on superinductances or a high-level of dissipation engineering. The former can require of the other of $\sim 200-300$ Josephson junctions to realize, which increases the probability of dielectric loss from two level systems. Whereas the latter has to be highly fine tuned to each qubit and can lead to new sources of error due to the various kinds of driving required to stabilize the computational states. Hence, the $Z_{2\beta}$ entanglemon stands as a scalable qubit with dual protection in superconducting circuits. Moreover, we are not aware of any proposal in the solid state literature (quantum dots for example) that would provide comparable protection as the $U(1)_\beta$ entanglemon. Given the very appreciable suppression of bit-flip errors in this model and the dominance of such errors in standard spin qubits, we hope our proposal could be utilized to achieve scalable protected qubits in semiconducting quantum dots. Finally, the architecture of the entanglemon and the protection from correlated (symmetric) noise as described above, makes both models attractive from the point of view of reducing the overhead for quantum error correction and reducing the occurences of correlated bit-flips - a key requirement for error correction.

While disorder in circuit parameters can complicate the spectrum of the entanglemon implementations, as long as this is small the lowest energetic states of such all to all generalized ferromagnetic models should still be accurately described by the coherent state description we present here for the logical states. We leave a detailed quantitative calculation of the above for future work.

\section{Discussion}
In this work, we have presented a cross-platform construction of a qubit utilizing entanglement. Entanglemons require only the presence of a coupled array of $\mathbb{C}$P(3) degrees of freedom and can hence be realized in a hardware-efficient way on a wide variety of quantum computing platforms.  The qubit relies on quantizing a collective mode of $\mathbb{C}$P(3) degrees of freedom and can be  well protected from depolarizing and dephasing errors.

More broadly, our work motivates the exploration of geometric non-linearities in compact phase-space manifolds, both for achieving fault-tolerance, for their inherent geometrical richness and the possibility of emergent weakly coupled degrees of freedom. Such an exploration of intrinsic geometric non-linearities, especially in the quantum context, has historically proven to be of great value in physics. For example, in high-energy physics a major breakthrough was Yang-Mills theory (where curvature arose due to through the non-abelian nature of the gauge theory and the notion of parallel transport in its lattice formulation), the discovery of asymptotic freedom in quantum Yang-Mills and the corresponding development of QCD \cite{yangmills,grossasymp,jaffe2006quantum}.
A little closer to the authors' interests perhaps is Haldane's pioneering work which introduced the non-linear sigma model and the geometry of semiclassical phase-space as a tool to describe the unusual properties of quantum spin chains \cite{Haldanenonlin}.

Several interesting directions for further  explorations suggest themselves. First, one could explore possibilities of realizing other compact phase-space manifolds and examining their noise-protection properties. While this might appear very theoretical, they have experimental promise. For example, graphene at charge neutrality realizes a Grassmanian manifold as its phase space --- Gr(2,4). Graphene quantum dots could potentially access this phase-space and it would be interesting to see the error-protection features of such Grassmanian spaces. Further, with the advent of moir\'e platforms, the potential to realize these higher-dimensional phase-spaces has also increased due to several pseudospin degrees of freedom present in such systems. Due to the great tunability in such platforms, one could envision realizing  modified version of the entanglemon.

Secondly, the natural next step would be  the development of universal quantum computing schemes and the physical protocols behind performing the required gates. This aspect would necessarily need to be  tailored to the specific hardware platform as would the specific protocols for initialization and measurement of the entanglemon. One attractive feature in this regard for the entanglemon is the existence of the symmetric and the antisymmetric subspace. One can exploit this feature to introduce a symmetric perturbation (say via a drive) which takes one out of the computational subspace, by tuning the drive frequency one can perform single qubit gates. Such approaches of utilizing non-computational states to perform gates in protected qubits is fairly standard by now. We hope future works address these issues in detail.

\section{Acknowledgments}
The authors thank S. Parameswaran, S. H. Simon and A. Chandran for valuable discussions and comments. This work was in part supported by the Deutsche Forschungsgemeinschaft under grants SFB 1143 (project-id 247310070) and the cluster of excellence ct.qmat (EXC 2147, project-id 390858490).

\appendix

\section{Primer on geometric quantization}
\label{subsec_geomquant}
Geometric quantization is a mathematical approach that allows one to smoothly connect classical and quantum mechanics. The main goal of the geometric quantization program is to construct a Hilbert space $\mathcal{H}_\hbar (\mathcal{M})$, depending on some measure of quantumness, say $\hbar$ or $S \approx 1/\hbar$, for a given symplectic manifold/phase-space $\mathcal{M}$. In our specific case, the symplectic manifold is a higher dimensional complex projective space $\mathbb{C}$P(3) and the quantization procedure is known as Berezin-Toeplitz quantization \cite{berezin1975quantization}.

This particular appendix contains a lot more than what is needed to understand the physics of the entanglemon construction, hence readers not interested in the idea of geometric quantization can skip this appendix and start from appendix B. Here we present a short primer on the main ideas of geometric quantization of complex projective spaces (developed in the mathematical community) in some generality and with some mathematical detail, as we believe it to be a very powerful framework with possible cross-field applications. Moreover, the mathematical details presented here are relatively lesser known to physicists and hence could serve as a nice starting point to explore the richness of geometric quantization. 

The main takeaways are how to construct a (quantum) Hilbert space from a complex projective space and how to obtain a dictionary between functions (symbols) over the complex projective space and operators in the corresponding (quantum) Hilbert space. In the semiclassical analysis (large-$d$ limit) presented in the main text, these correspondences simplify greatly. This appendix allows a deeper understanding of the  navigation between the classical and quantum Hamiltonians presented for the simple models in sec. \ref{sec_simplemodels}.

\subsection{K\"ahler metric on projective spaces}

The projective space $\mathbb{C}$P($n$) is the set of complex lines in $\mathbb{C}^{n+1}$. Hence, we have to identify $(z_0,z_1,\dots,z_n)$ and $(\lambda z_0,\lambda z_1,\dots , \lambda z_n)$ for arbitrary complex $\lambda \neq 0$ (defining property of $\mathbb{C}$P($n$)). For a complex manifold description of such a projective space we need to cover it with charts (patches). A common choice is based on the open subsets $U_i = {(z_0:z_1:\dots:z_n) \in \mathbb{C}\text{P}(n); z_i \neq 0}$.

If $z_i \neq 0$ the ray through ($z_0,z_1,\dots,z_n$) is characterized uniquely by $(z_0/z_i,\dots,z_n/z_i) \equiv (w_0^{(i)},\dots,w^{(i)}_{i-1},1,w^{(i)}_{(i+1)},\dots,w_n^{(i)})$. So $(w_0^{(i)},\dots,w^{(i)}_{i-1},w^{(i)}_{(i+1)},\dots,w_n^{(i)})$ can be considered to be independent coordinates on $U_i$. On $U_i \cap U_j$, we get two sets of coordinates \{$w_k^{(i)},k \neq i$\} and \{$w_k^{(j)}, k \neq j$\}. The transformation between these is given by
\begin{equation}
\begin{aligned}
    w_k^{(i)} &= \frac{z_k}{z_i} = \frac{z_k/z_j}{z_i/z_j} = \frac{w_k^{(j)}}{w_i^{(j)}}; \,k \neq i,j \\
    w_j^{(i)} &= \frac{z_j}{z_i} = \frac{1}{w_i^{(j)}}
\end{aligned}
\end{equation}
Hence, the map from $w^{(j)}$ to $w^{(i)}$ is holomorphic since $w_i^{(j)} \neq 0$ in $U_i \cap U_j$. Note that the complement of $U_i$ in $\mathbb{C}$P($n$) is homeomorphic to $\mathbb{C}$P($n-1$), and hence is a set of measure zero. So, in many applications, for example for writing the completeness relations for generalized coherent states, we can use a single chart. This will be important later on.

Now, we know that $\mathbb{C}^{n+1}$ is endowed with a natural metric which is invariant under dilations $(z_0,z_1,\dots,z_n) \rightarrow (\lambda z_0,\lambda z_1,\dots,\lambda z_n)$. Let's take two tangent vectors $(u_0,\dots,u_d)$ and $(v_0,\dots,v_n)$ at $(z_0,\dots,z_n)$, their hermitian product is defined as $\frac{<u|v>}{<z|z>}-\frac{<u|z><z|v>}{<z|z>^2}$ which is invariant under $u_i \rightarrow \lambda u_i$ and similar transformations for $v_i,z_i$. The second term in the product is to ensure that tangent vectors proportional to $z$ are orthogonal to any tangent vector. This allows the construction of a metric on $\mathbb{C}$P($n$):
\begin{equation}
    g(u,v) = \dfrac{\sum_{j \neq i} \bar{u}_j^{(i)}v_j^{(i)}}{1+\sum_{j \neq i}|w_j^{(i)}|^2} - \dfrac{\sum_{j \neq i} \sum_{k \neq i} w_j^{(i)} \bar{w}_k^{(i)}\bar{u}_j^{(i)}v_k^{(i)}}{(1+\sum_{j \neq i}|w_j^{(i)}|^2)^2}
\end{equation}

Using Berezin's notations, the above metric can be written as a Hermitian metric
\begin{equation}
\begin{split}
    ds^2 = \sum_{j \neq,k \neq i} g_{j\bar{k}}^{(i)}dw_j^{(i)}d\bar{w}_k^{(i)}\\
    g_{j\bar{k}}^{(i)} = \dfrac{\delta_{jk}}{1+<w^{i}|w^{i}>} - \dfrac{\bar{w_j}^{(i)}w_k^{(i)}}{(1+<w^{i}|w^{i}>)^2}
\end{split}
\end{equation}
where the superscript (i) indicates that this expression of the metric involves the coordinates $w_j^{(i)}$ adapted to the chart $U_i$.

From the above metric, there is a built-in real antisymmetric form $\omega$ which is obtained by taking the imaginary part of the hermitian form.
\begin{equation}
    Im (\sum_{j,\bar{k} \neq i} (g_{j\bar{k}}^{(i)}u_j^{(i)}\bar{v}_k^{(i)})) = \frac{1}{2i} \sum_{j,k \neq i} g_{j \bar{k}}^{(i)}(u_j^{(i)}\bar{v}_k^{(i)} - v_j^{(i)}\bar{u}_k^{(i)})
\end{equation}
We thus define the two-form
\begin{equation}
    \omega = \frac{1}{2i} \sum_{j,k \neq i} g_{j\bar{k}}^{(i)}(dw_j^{(i)} \wedge d\bar{w}_k^{(i)})
\end{equation}
Crucially, this form $\omega$ allows us to view $\mathbb{C}$P($n$) as a classical phase-space for Hamiltonian dynamics if $\omega$ is closed. Such a closed condition in this situation implies
\begin{equation}
    \dfrac{\partial g_{j\bar{k}}^{(i)}}{\partial w_{l}^{(i)}} = \dfrac{\partial g_{l\bar{k}}^{(i)}}{\partial w_{j}^{(i)}}; \quad \dfrac{\partial g_{j\bar{k}}^{(i)}}{\partial \bar{w}_{k}^{(i)}} = \dfrac{\partial g_{j\bar{l}}^{(i)}}{\partial \bar{w}_{k}^{(i)}}
\end{equation}
These conditions are satisfied if the metric derives from a K\"ahler potential, i.e, we have functions $\varphi^{(i)}(w^{(i)},\bar{w}^{(i)})$ (in $U_i$) such that
\begin{equation}
    g_{j\bar{k}}^{(i)} = \dfrac{\partial^2 \varphi^{(i)}}{\partial w_j^{(i)} \partial \bar{w}_k^{(i))}}
    \label{Kahpot}
\end{equation}
The existence of these functions can be checked and there form can be derived, however, we do not present those calculations here. We simply state the result, the K\"ahler potential has the form
\begin{equation}
    \varphi^{(i)} = \text{log}(1+\langle \bm{w}^{(i)}| \bm{w}^{(i)} \rangle)
\end{equation}
where $\bm{w^{(i)}} = \{ w_1^{(i)},\dots,w_n^{(i)}\}$
\subsection{Inner product on quantum states}
In Berezin-Toeplitz quantization, the quantum Hilbert space is realized as a space of holomorphic functions of $w_1^{(0)},\dots,w_n^{(0)}$. Knowing the underlying phase-space manifold and its metric, the next step is to define an inner product in such a space of functions. 
Following ideas proposed by Berezin \cite{berezin1975quantization}, the inner product can be written as
\begin{equation}
\begin{split}
    \langle f|g \rangle = c(d) \int \overline{f(\bm{w})}g(\bm{w})
    \text{exp}(-d \, log(1+\langle \bm{w}|\bm{w} \rangle))\\
    d \mu(\bm{w})
\end{split}
\end{equation}
where $w_i = w_i^{(0)}$ and $\bm{w} = \{w_1,\dots,w_n\}$ for notational convenience. The K\"ahler potential $\varphi^{(0)}$ appears in the exponential factor. The final step is then to specify the measure $d\mu$. For a K\"ahler manifold, a natural choice is to take
\begin{equation}
d\mu \propto \omega \wedge .. \wedge \omega ( \text{n times}) \propto \rm{Det} [g^{(0)}_{j\bar{k}}] \prod_{i=1}^n dw_i d\bar{w}_i; 
\label{Kahlmet}
\end{equation}
where $1 \leq j,k \leq n$ and 
\begin{equation}
\begin{aligned}
    \rm{Det} \big[ g_{j\bar{k}}^{(0)}\big] &= \dfrac{1}{(1+\langle w | w \rangle)^n}\, \rm{Det}[M_{j \bar{k}}] \\
    M_{j \bar{k}} &= \delta_{jk} - \dfrac{\bar{w}_j w_k}{1+ \langle w | w \rangle}
\end{aligned}
\end{equation}
$M$ has two eigenvalues, $1/(1+\langle w| w \rangle)$ and $1$. The first one is non-degenerate and corresponds to the eigenvector proportional to $(w_1,...,w_d)$ whereas the latter is $n-1$ degenerate and is associated to the eigenspace orthogonal to $(w_1,...,w_d)$. Hence, we get 
\begin{equation}
    \rm{Det}[g_{j \bar{k}}^{(0)}] = \dfrac{1}{(1+\langle w | w \rangle )^{n+1}}
\end{equation}
which then gives us our measure by substituting the above in eq. (\ref{Kahlmet}).

For reasons we briefly highlighted in the main text and will also come across here soon, possible values of the Planck's constant $h$ are quantized: $h = 1/d$ where $d \in \mathbb{Z}^+$. For $n=2$ we have the usual SU(2) spin and $d = 2S$ where $S$ is a half integer. Our Hilbert space $\mathcal{H}_d$ is then defined with the following inner product
\begin{equation}
    \langle f|g\rangle_d = c(d) \int \dfrac{\overline{f(w)}g(w)}{(1+\langle w|w \rangle)^{n+1+d}} \prod_{j=1}^d dw_j d\bar{w}_j
    \label{inprod}
\end{equation}

\subsection{Quantum Hilbert space $\mathcal{H}_d$ and classical limit}
\label{quanthilb}
A holomorphic function in chart $U_0$ can be represented by a power series
\begin{equation}
f(w) = \sum_{a_1,...,a_j \in N} c(a_1,...,a_n) w_1^{a_1}...w_n^{a_n}
\end{equation}
where $f$ is normalizable if and only if its total degree is finite and less than equal to $d$ (as mentioned in the last section, we hide the (0) in the superscript of the $w$ to avoid clutter). If $a_1+\dots+a_n \leq d$, it will prove convenient to introduce $a_0 \geq 0$ such that $a_0+a_1+\dots+a_n = d$. From this condition, using standard combinatorics arguments one can obtain the Hilbert space dimension
\begin{equation}
    \text{dim}\,\mathcal{H}_d = \frac{(n+d)!}{n!\,d!}
\end{equation}

To see how one recovers the classical limit from such a quantum Hilbert space, we need to consider how $f(w)$ defined on chart $U_0$ can be extended to a system of functions $f^{(i)}$ on charts $U_i$. For $f \in \mathcal{H}_d$ and $\bm{a} = \{a_1,...,a_n \}$, we have $f(w) = \sum_{a_1+..+a_n \leq d} c(\bm{a})w_1^{a_1}...w_n^{a_n}$. On chart $U_0$, we set $f^{(0)}(w) = \frac{1}{z_0^d} \sum_{a_0+..+a_n = d} c(\bm{a})z_0^{a_o}z_1^{a_1}...z_n^{a_n}$. This suggests the definition
\begin{equation}
    f^{(i)}(z_0:,..:,z_n) = \frac{1}{z_i^d}\sum_{a_0+...+a_n = d} c(\bm{a})z_0^{a_0}...z_n^{a_n}
\end{equation}
which is constant on the ray through $(z_0,...,z_n)$ because both numerator and denominator are of degree $d$. This gives us the relation
\begin{equation}
    f^{(i)} = \bigg(\frac{z_j}{z_i}\bigg)^d f^{(j)} \quad \text{on} \quad U_i \cap U_j
\end{equation}
These relations show that the collection $f^{(i)}$ of local holomorphic functions on $U_i$ define a global section of a line bundle whose transition functions are
\begin{equation}
    h_{ij} = (\frac{z_j}{z_i})^d \quad \text{on} \quad U_i \cap U_j
\end{equation}
These transition functions obey the usual relations $h_{ij} = h_{ji}^{-1}$ on $U_i \cap U_j$ and $h_{ij}h_{jk} = h_{ik}$ on $U_i \cap U_j \cap U_k$. Moreover the above relation shows that the line bundle at a given $d$ is obtained from the fundamental one at $d=1$ by taking its $d$-th tensor product. This is analogous to constructing a spin $S$ by taking the symmetrized tensor product of $2S$ spin-$1/2$ Hilbert spaces. Hence, taking $d \rightarrow \infty$ corresponds to a classical limit of this quantum Hilbert space.

\begin{table*}[t!]
\centering
\begin{tabular}{|c|c|}
  \hline
  Symbol & Operator  \\ 
  \hline
  $\dfrac{v_i}{1+ \langle v | v \rangle}$  & $\frac{1}{n+d+1} \hat{b}_0 \hat{b}_i^{\dagger}$  \\ 
  $\dfrac{\bar{v}_i}{1+ \langle v | v \rangle}$  & $\frac{1}{n+d+1} \hat{b}_i \hat{b}_0^{\dagger}$\\ 
  $\dfrac{v_i \bar{v}_j}{1+ \langle v | v \rangle}$  & $\frac{1}{n+d+1} \hat{b}_j \hat{b}_i^{\dagger}$ \\
  $\dfrac{v_i v_j}{1+ \langle v | v \rangle}$  & $\frac{1}{n+d+2}\frac{1}{n+d+1} \hat{b}^2_0 \hat{b}_i^{\dagger}\hat{b}_j^{\dagger}$\\
  $\dfrac{\bar{v}_i \bar{v}_j}{\big(1+ \langle v | v \rangle \big)^2}$  & $\frac{1}{n+d+2}\frac{1}{n+d+1} \hat{b}_i\hat{b}_j(\hat{b}_0^{\dagger})^2$\\
  \hline
\end{tabular}
\caption{The dictionary for converting from contravariant symbols to operators under the scheme of geometric quantization on the $\mathbb{C}$P($n-1$) manifold, see appendix. \ref{subsec_symtoop}}
\label{tabop}
\end{table*}

\subsection{Orthonormal basis and reproducing kernel}
Consider the state $\ket{\bm{a}}$ associated to the monomial $w_1^{a_1}\dots w_n^{a_n}$ (in $U_0$). From the definition of the inner product in eq. (\ref{inprod}), we get that $\langle \bm{a}|\bm{b} \rangle = 0$ when $\bm{a} \neq \bm{b}$, since then the angular integral vanishes (if we use polar co-ordinates for each $w$). Then,
\begin{equation}
\begin{split}
    \langle \bm{a}|\bm{a} \rangle = c(d) (2\pi)^n \int dr_1 ... \int dr_n \,\dfrac{r_1^{2a_1+1}\dots r_n^{2a_n+1}}{(1+n_1^2+..+n_d^2)^{n+1+d}}\\
    = c(d) \pi^n \int dx_1 ... \int dx_n \,\dfrac{x_1^{a_1}\dots x_n^{a_n}}{(1+x_1+..+x_n)^{n+1+d}}
\end{split}
\end{equation}
To simplify the above expression consider the integrals
\begin{equation}
\begin{split}
    I_{p,q}(a) \equiv \int^\infty_a \frac{x^p}{(a+x)^{q+1}}\,dx = \frac{1}{a^{q-p}}I_{p,q}(1); \quad p<q \\
    = \dfrac{p!\,(q-p-1)!}{q!} \frac{1}{a^{q-p}}
\end{split}
\end{equation}
Using the above and iteratively going through every integral we get the important relation
\begin{equation}
    \langle\bm{a}|\bm{a} \rangle = \dfrac{c(d)\pi^n}{(d+n)!}\,a_0!\,a_1!...a_n!
\end{equation}
One can simplify the above expression by choosing $c(d) = (d+n)!/\pi^n$. We can then see that $\mathcal{H}_d$ is isomorphic to the subspace in the Fock space of $n+1$ bosonic
modes subjected to the constraint:
\begin{equation}
    \sum_{i= 0}^{n} b^{\dagger}_i b_i = d
\end{equation}
The above is a generalization of the Schwinger boson representation of SU(2) spins with $d = 2S$. The state $\ket{\bm{a}}$ is in correspondence with $(b_0^{\dagger a_o}b_1^{\dagger a_1}...b_n^{\dagger a_n} \ket{0}$ (they have the same norm). An orthonormal basis is obtained as:
\begin{equation}
    f_{\bm{a}}(w_1,...,w_n) = \dfrac{1}{\sqrt{a_0!a_1!...a_n!}}\, w_1^{a_1}...w_n^{a_n}
    \label{orthbasis}
\end{equation}
One can also define an object called the reproducing kernel which will come in handy in the next subsection
\begin{equation}
\begin{split}
    L_d(w,\bar{w}) = \sum_{a_0+..+a_n = d} \dfrac{1}{a_0!..a_n!}\,|w_1|^{2a_1}...|w_n|^{2a_n} \\
    = \frac{1}{d!}\,(1+\langle w|w \rangle )^d
\end{split}
\end{equation}
This kernel is directly related to the K\"ahler potential (described earlier in eq. \ref{Kahpot}) as
\begin{equation}
    L_d(w,\bar{w}) = \frac{1}{d!} \exp{(d \varphi(w,\bar{w}))}
\end{equation}

\subsection{From contravariant symbols to operators}
\label{subsec_symtoop}
Now we come to the main step of Toeplitz quantization which is to obtain quantum operators from classical functions (symbols). Let us consider a function $A(v,\bar{v})$ where $v,\bar{v} \in \mathbb{C}\rm{P}(n)$. We want to find the quantized operators of this function in the (quantum) Hilbert space $\mathcal{H}_d$ (see appendix \ref{quanthilb}). The first step to do so is define generalized coherent states on $\mathbb{C}$P($n$). For any $\psi \in \mathcal{H}_d$ we have the identity, involving the reproducing kernel:
\begin{equation}
    \psi(w) = c(d) \int \dfrac{\prod_{j=1}^n dv_j d\bar{v}_j}{(1+\langle v|v \rangle)^{d+1+n}}L_d(w,\bar{v}) \psi(v)
\end{equation}
Coherent states are then defined by $CS_{\overline{v}}(w) \equiv L_d(w,\bar{v})$. Using that $L_d(w,\bar{v}) = \overline{CS_{\bar{w}}(v)}$ the above equation becomes
\begin{equation}
    \psi(w) = \langle CS_{\bar{w}} | \psi \rangle
    \label{gen_def_coh_state}
\end{equation}
which is a defining property of a generalized coherent state in the Berezin-Toeplitz quantization scheme. Formally, $\ket{CS_{\bar{w}}}$ is orthogonal to the hyperplane of holomorphic functions vanishing at $w$, which is a simple way of defining a state localized around $w$. Using this we can get a resolution of the identity operator as
\begin{equation}
    \mathbb{1} = c(d) \int \dfrac{\prod_{j=1}^n dv_j d\bar{v}_j}{(1+\langle v|v \rangle)^{d+1+n}} \ket{CS_{\bar{v}}}\bra{CS_{\bar{v}}}
    \label{idenres}
\end{equation}
From~(\ref{gen_def_coh_state}), $\langle CS_{\bar{v}}|CS_{\bar{v}} \rangle = L_d(v,\bar{v}) = (1+\langle v|v \rangle)^d/d!$ which then gives us the important identity 
\begin{equation}
    \mathbb{1} = \frac{c(d)}{d!} \int \dfrac{\prod_{j=1}^d dv_j d\bar{v}_j}{(1+\langle v|v \rangle)^{n+1}} \dfrac{\ket{CS_{\bar{v}}}\bra{CS_{\bar{v}}}}{\langle CS_{\bar{v}}|CS_{\bar{v}} \rangle}
\end{equation}
From the above we see that the integrand does not depend on the "Planck constant" $1/d$, the only dependence is through the prefactor $\frac{c(d)}{d!} = \frac{(n+d)!}{\pi^n d!}$.  In the semi-classical limit ($d \rightarrow \infty$), this prefactor is equivalent to $(d/\pi)^n$ which is consistent with the intuitive notion that a quantum state is associated with a phase-space volume of order $(\frac{1}{d})^n$.

For a function $A(v,\bar{v})$ (contravariant symbol), we get the operator $\hat{A}$ defined by the generalization of eq. \ref{idenres}:
\begin{equation}
 \hat{A} = c(d) \int \dfrac{\prod_{j=1}^n dv_j d\bar{v}_j}{(1+\langle v|v \rangle)^{d+1+n}}A(v,\bar{v})\ket{CS_{\bar{v}}}\bra{CS_{\bar{v}}}   
\end{equation}
In practice however, one obtains $\hat{A}$ by computing its matrix elements of $\hat{A}$ in the orthonormal basis $\ket{f_{\bm{a}}}_d$ where $a_0+..+a_n = d$ as in eq. (\ref{orthbasis})
\begin{equation}
    \langle f_{\bm{b}} | \hat{A} | f_{\bm{a}} \rangle = c(d) \int \dfrac{\prod_{j=1}^d dv_j d\bar{v}_j}{(1+\langle v|v \rangle)^{n+1+d}} A(v,\bar{v})\overline{f_{\bm{b}}(v)}f_{\bm{a}}(v)
\end{equation}
One can use this relation to get the correspondence for functions and operators, some examples of which are shown in table \ref{tabop}.

\section{Berry phase considerations}
We want to calculate the Berry phase associated with a cyclic ($2\pi$) rotation of $\beta$. Consider an infinitesimal variation $\beta \rightarrow \beta + \delta \beta$ which results in change $\delta \Psi$ of the $\mathbb{C}$P(3) spinor. In the coherent state notation in sec. \ref{subsec_quantcoherent} such a variation yields a $\delta \hat{b}^{\dagger} (\Psi)$ and hence the resulting change in the coherent state $\delta \ket{CS_d (\Psi)}$. Now, we have the relation
\begin{equation}
    \delta \ket{CS_d (\Psi)} = \sqrt{\frac{d}{(d-1)!}}\, \delta \hat{b}^{\dagger}(\Psi) \, \hat{b}^{\dagger}(\Psi)^{d-1} \ket{0}
\end{equation}
Further, we can write
\begin{equation}
    \begin{aligned}
        &\langle 0| \hat{b}^d(\Psi) \delta b^{\dagger}(\Psi) b^{\dagger}(\Psi)^{d-1} | 0 \rangle \\
        &= d\, \big[\hat{b}(\Psi),\delta b^{\dagger}(\Psi)\big] \langle 0|b(\Psi)^{d-1}b^{\dagger} (\Psi)^{d-1} | 0 \rangle \\
        &= d(d-1)!\,\big[ \hat{b}(\Psi),\delta \hat{b}^{\dagger} (\Psi)\big]
    \end{aligned}
\end{equation}
using which we obtain the relation
\begin{equation}
    \langle CS_d (\Psi)|\delta CS_d (\Psi) \rangle = \frac{d}{2}\dfrac{\langle \Psi | \delta \Psi \rangle - \langle \Psi | \delta \Psi \rangle}{\langle \Psi | \Psi \rangle}
\end{equation}
which then boils down to a standard Berry phase associated with the solid angle subtended in the entanglement Bloch sphere.
\section{Classical aspects of the $U(1)^\beta$ symmetry}
\label{App_sec_beta_translations}

To examine effects of order by disorder we first calculate the symplectic form on $\mathbb{C}$P(3) using the natural $(\alpha,\beta,\theta_{\rm{S}},\varphi_{\rm{S}},\theta_{\rm{P}},\varphi_{\rm{P}})$ variables.
These are defined by:
\begin{equation}
    \ket{\psi} = \text{cos}\frac{\alpha}{2}\ket{\varphi_S} \otimes \ket{\varphi_P} + \text{sin}\frac{\alpha}{2}e^{i\beta}\ket{\chi (\theta_S,\varphi_S)}\otimes \ket{\chi_P(\theta_P,\varphi_P)}
\end{equation}
The Berry connection (1-form) is then given by
\begin{equation}
\begin{aligned}
    \mathcal{A} &= \frac{1}{2i}\bigg( \langle \psi | d\psi \rangle - \langle d\psi | \psi \rangle \bigg) = \\
    & \text{sin}^2 \frac{\alpha}{2}\,d\beta + \cos{\alpha}\, \bigg( \text{sin}^2 \frac{\theta_S}{2}\, d \varphi_S + \text{sin}^2 \frac{\theta_S}{2}\, d \varphi_P \bigg)
\end{aligned}
\end{equation}

From the above description we can then derive the symplectic form

\begin{equation}
\begin{split}
    \omega = d \mathcal{A} = \frac{1}{2}\sin{\alpha}\, d\alpha \wedge d\beta + \mbox{} \\ \frac{1}{2}\cos{\alpha}\, (\sin{\theta_S}\, d\theta_S \wedge d\varphi_S + \sin{\theta_P}\, d\theta_P \wedge d \varphi_P) - \mbox{} \\
    \sin{\alpha}\, (\text{sin}^2 \frac{\theta_S}{2}\,d\alpha \wedge d\varphi_S + \text{sin}^2 \frac{\theta_P}{2}\,d\alpha \wedge d\varphi_P)
\end{split}  
\label{express_omega}
\end{equation}

We see that the symplectic form above $\omega$ is invariant under $U(1)^\beta$ transformations, i.e under $\beta$ translations. This shows that, at least locally, these $\beta$ translations can be obtained as the Hamiltonian flow associated to a generating function $G(\alpha,\beta,\theta_{\rm{S}},\varphi_{\rm{S}},\theta_{\rm{P}},\varphi_{\rm{P}})$ on the $\mathbb{C}$P(3) manifold. At the quantum level, such transformations correspond to unitary operators acting on the Schwinger boson Hilbert space. 
As a result order by disorder mechanisms cannot break the underlying 
$U(1)^\beta$ symmetry down to $\mathbb{Z}_2^\beta$.
 
Let us also evaluate the generating function $G$.
This function is requested to produce a Hamiltonian velocity field
$V_j$ ($j \in \{\alpha,\beta,\theta_{\rm{S}},\varphi_{\rm{S}},
\theta_{\rm{P}},\varphi_{\rm{P}}\}$) such that $V_{\beta}=1$
and $V_j =0$ for $j \neq \beta$. The Hamilton's equations of motion
read:
\begin{equation}
    \sum_{j} \omega_{ij} V_j = \partial_i G 
\end{equation}
for any $i \in \{\alpha,\beta,\theta_{\rm{S}},\varphi_{\rm{S}},
\theta_{\rm{P}},\varphi_{\rm{P}}\}$.
From the above expression~(\ref{express_omega}) for the symplectic form,
we get that $\partial_i G=0$, except for $i=\alpha$ for which we have
$\partial_{\alpha} G=(1/2)\sin{\alpha}$, so we may choose
$G=-\frac{1}{2}\cos{\alpha}$.

\section{2 site model with exchange interactions}
Here we provide details on the 2 site (dot) model with a spin and pseudospin-1/2 degrees of freedom in each dot and an antiferromagnetic exchange coupling between the two dots. Such a model, as mentioned in the main text reduces the $U(1)^\beta$ symmetry associated with  $(\beta^a,\beta^b) \rightarrow (\beta^a+\theta,\beta^b+\theta)$ to $\mathbb{Z}^\beta_2$, i.e invariance under $(\beta^a,\beta^b) \rightarrow (\beta^a+\pi,\beta^b+\pi)$. However, the said $\mathbb{Z}^\beta_2$ is still part of a larger antisymmetric $U(1)^\beta$ associated with  $(\beta^a,\beta^b) \rightarrow (\beta^a+\theta,\beta^b-\theta)$.

We introduce two sites ($a$ and $b$) with the same Hamiltonian as in model $U(1)^{\beta}$ and a coupling between the two sites.
\begin{equation}
    H_{1B} = H_{0a} + H_{0b}  + H_{\rm{int}}
    \label{2smod}
\end{equation}
Generic interaction term that could enter the coupling has the form
\begin{equation}
    H_{\rm{int}} = \lambda( |\langle \bm{\Psi}^a|\bm{\Psi}^b \rangle |^2)
    \label{2smodint}
\end{equation}
We consider antiferromagnetic coupling between the two sites, i.e. $\lambda > 0$. 
$H_{0a}$ and $H_{0b}$ do not pickup any terms sensitive to $\beta$, hence for the initial step we focus on $H_{\rm{int}}$. We write
\begin{equation}
\begin{split}
    \langle \bm{\Psi}^a|\bm{\Psi}^b \rangle  = \rm{cos}\frac{\alpha^a}{2}\rm{cos}\frac{\alpha^b}{2}\langle \varphi^a|\varphi^b \rangle + \\ e^{i(\beta_b-\beta_a)}\rm{sin}\frac{\alpha^a}{2}\rm{sin}\frac{\alpha^b}{2}\langle \chi^a|\chi^b \rangle +\\
   e^{i\beta_b} \rm{cos}\frac{\alpha^a}{2}\rm{sin}\frac{\alpha^b}{2} \langle \varphi^a | \chi^b \rangle + e^{-i\beta_a}\rm{cos}\frac{\alpha^b}{2}\rm{sin}\frac{\alpha^a}{2}\langle \chi^a | \varphi^b \rangle
\end{split}
\end{equation}
 The sum of the two terms above is invariant under $(\beta_a,\beta_b) \rightarrow (\beta_a+\theta,\beta_b+\theta)$ for any $\theta$. We want to find conditions such that this $U (1)_\beta$ symmetry is reduced to a $\mathbb{Z}^\beta_2$. One can show that the above requirement is reduced to finding a condition on the parameters on the two sites such that 
 $\langle \varphi^a | \varphi^b \rangle = \langle \chi^a | \chi^b \rangle = 0$ \textit{and} that $\langle \varphi^a (\chi^a) | \chi^b (\varphi^b) \rangle \neq 0$. These conditions imply that all single phase ($e^{i\beta^{a(b)}}$) or phase-difference ($e^{i(\beta^a - \beta^b)}$) terms vanish in the expansion of the $|\langle \Psi_a|\Psi_b \rangle |^2$ term, and the phase sum terms ($e^{i(\beta^a+\beta^b)}$) are the only $\beta$-dependent terms. As a result the invariance is reduced to only $\theta = \pi$. Therefore, $U(1)^{\beta}$ is broken to $\mathbb{Z}^\beta_2$. Hence, we need to find solutions to $ e^{i\beta_b} \langle \varphi^a | \chi^b \rangle + e^{-i\beta_a} \langle \chi^a | \varphi^b \rangle = 0$ satisfying $\langle \varphi^a (\chi^a) | \chi^b (\varphi^b) \rangle \neq 0$.
Close examination of the above condition reveals that such conditions are satisfied for
\begin{equation}
\begin{split}
    \ket{\varphi^a} = \ket{\chi^b}; \ket{\varphi^b} = \ket{\chi^a}; \alpha^a,\alpha^b \neq 0,\pi \\
    \beta^a + \beta^b = (2n+1)\pi; n \in \mathbb{Z}
    \end{split}
    \label{1bcond}
\end{equation}

Whether there exist states satisfying the above constraints, which are also ground states of the interaction term is the next pertinent question. Note that unlike standard SU(2) spins, because of the larger dimensionality of $\mathbb{C}$P(3) space(6-dimensional), there exists a large family of generalized antiferromagnetic ground states. More specifically for every $\Psi^a \in$ $\mathbb{C}$P(3) there is a 
$\mathbb{C}$P(2) manifold of states satisfying $\langle \Psi^a | \Psi^b \rangle = 0 $.  Moreover, the two possible easy-axis ground states of $H_0$, as shown in eq. (\ref{eags}) also satisfy this criteria. Hence, if only the first term is present, the natural ground state of $H_{1B}$ is each site having the form of one of the two easy-axis states, i.e spin and pseudospin entangled on each site, spins aligned and pseudospins anti-aligned. The full U(1) symmetry of the $H_0$ ground states is retained. The last two terms in the interaction in eq. (\ref{2smodint}) break this large degeneracy of the $\mathbb{C}$P(2) manifold, favouring states with opposite spin \textit{and} pseudospin on the two sites.

Let us examine the nature of the spinors for such anti-aligned states. We know that $\ket{\varphi^{a(b)}} = \ket{\varphi^{a(b)}_S} \otimes \ket{\varphi^{a(b)}_P}$ and $\ket{\chi^{a(b)}} = \ket{\chi^{a(b)}_S} \otimes \ket{\chi^{a(b)}_P}$. Expressing this in terms of angles in the Bloch sphere,we get
\begin{equation}
\begin{split}
    \ket{\varphi^{i}} = \begin{pmatrix}
        \rm{cos}\frac{\theta^{i}_S}{2} \rm{cos} \frac{\theta^{i}_P}{2} \\
        e^{i\varphi^i_P}\rm{cos}\frac{\theta^{i}_S}{2} \rm{sin} \frac{\theta^{i}_P}{2} \\
        e^{i \varphi^i_S}\rm{sin}\frac{\theta^{i}_S}{2} \rm{cos} \frac{\theta^{i}_P}{2} \\
        e^{i (\varphi^i_S+\varphi^i_P)}\rm{sin}\frac{\theta^{i}_S}{2} \rm{sin} \frac{\theta^{i}_P}{2}
    \end{pmatrix} \\
     \ket{\chi^{i}} = \begin{pmatrix}
       e^{-i (\varphi^i_S+\varphi^i_P)}\rm{sin}\frac{\theta^{i}_S}{2} \rm{sin}\frac{\theta^{i}_P}{2}\\
        -e^{-i\varphi^i_S}\rm{sin}\frac{\theta^{i}_S}{2} \rm{cos} \frac{\theta^{i}_P}{2} \\
        -e^{-i\varphi^i_P}\rm{cos}\frac{\theta^{i}_S}{2} \rm{sin} \frac{\theta^{i}_P}{2} \\
        \rm{cos}\frac{\theta^{i}_S}{2} \rm{cos} \frac{\theta^{i}_P}{2} \\
    \end{pmatrix}
\end{split}
\end{equation}
where $i = a,b$. For anti-aligned easy-axis states, we have $\varphi_{S(P)}^{a(b)} = 0$. Moreover, we have $\theta^a_{S(P)} = 0$ and $\theta^b_{S(P)} = \pi$. Hence subsituting these into the above expression we get

\begin{equation}
\begin{split}
    \ket{\varphi^{a}} =  \begin{pmatrix}
        1\\
        0 \\
        0\\
        0
    \end{pmatrix}; \ket{\varphi^{b}} =  \begin{pmatrix}
        0\\
        0 \\
        0\\
        1
    \end{pmatrix};
     \ket{\chi^{a}} =  \begin{pmatrix}
       0\\
        0\\
        0\\
        1\\
    \end{pmatrix};  \ket{\chi^{b}} =\begin{pmatrix}
        1\\
        0 \\
        0\\
        0
    \end{pmatrix}
\end{split}
\label{antspin}
\end{equation}
Hence, we see that such anti-aligned spin and pseudospins satisfy the first condition in eq. \ref{1bcond}, since $\ket{\varphi^{a(b)}} = \ket{\chi^{b(a)}}$. Further from eq. \ref{1bcond} we also see, that anti-alignment is not a sufficient condition, the spins and pseudospins must also be entangled since we require $\alpha \neq 0,\pi$. Can such anti-aligned entangled configurations be unique ground states of $H_{1B}$?

The only term that presents a hurdle for the realization is the $\Delta S^z$ term. Such a term comes from an external magnetic field and might be unavoidable in many hardware realizations (see next sections). This term favours the alignement of the spins along the direction of the magnetic field and hence will cause the spins on both sites to point along the same direction. However, the presence of such a term in the energy functional is imperative, since without it, there is a large degeneracy of the ground state manifold which contains all maximally entangled states. This means that the spins and pseudospins can point along any direction, on both sites, there is no energetic preference, since due to maximal entanglement on each site, the magnitude of the spin and pseudospin vectors is zero.

To circumvent this problem we introduce different coupling constants for such a $S^z$ term for the two sites. More, specifically we impose $\Delta^a = -\Delta^b$, hence the term $H_{0a}+H_{0b}$ in eq. \ref{2smod}. Hence, the large degeneracy of the ground state manifold of the interaction term is broken by the single-site terms. Moreover, note that the single-site terms (taking $\Delta^a = -\Delta^b$) select exactly the ground state configuration with 
\begin{equation}
\begin{split}
    \ket{\Psi^a} = \text{cos}\frac{\alpha}{2}\ket{\varphi^a} \pm e^{i\beta^a} \text{sin}\frac{\alpha}{2} \ket{\chi^a}\\
    \ket{\Psi^b} = \text{cos}\frac{\alpha}{2} \ket{\varphi^b}\pm e^{i\beta^b} \text{sin}\frac{\alpha}{2} \ket{\chi^b})
    \label{gstat1b}
\end{split}
\end{equation}
where $\ket{\varphi^{a(b)}}$,$\ket{\chi^{a(b)}}$ are given in eq. \ref{antspin}, $\text{cos}\alpha = \Delta^a/(2u_z) = -\Delta^b/(2u_z) $ and the $\pm$ is due to the $\mathbb{Z}^\beta_2$. 

Hence, we have shown that in the regime $u_p > u_z>\Delta^a/2>0$ and $\lambda > 0$, the two-site model in Eq. \ref{2smod} and \ref{2smodint} reduces the $U(1)^\beta$ symmetry, $(\beta^a,\beta^b) \rightarrow (\beta^a+\theta,\beta^b+\theta)$ to $\mathbb{Z}^\beta_2$ with ground states given in eq. \ref{gstat1b}, where the ground state manifold is invariant only under $(\beta^a,\beta^b) \rightarrow (\beta^a+\pi,\beta^b+\pi)$. However, as we see in eq. \ref{1bcond}, there is a remnant anti-symmetric U(1) associated with $(\beta^a,\beta^b) \rightarrow (\beta^a+\theta,\beta^b-\theta)$. 

\section{Gap analysis for entanglemon implementations}

In this appendix we present simple calculations justifying the gap between the logical states and other symmetric/non-symmetric states as mentioned in sec. VI, for the implementations of the two entanglemon models as described in sec. \ref{sec_implementations}. 

Let us first consider the $Z_{2\beta}$ case, in the implementations for which we saw in sec. \ref{sec_implementations} that we get an effective quantum Ising model of the form $-J \sum \tau_i \tau_j$. Now if we take $n$ copies with $\tau_i = +1$ and $d-n$ copies with $\tau_i = -1$ this defines a degenerate energy eigenspace with $E_n = E_0 + 2n(d-n)$ with degenerancy $d!/(n!(d-n)!)$. For each $n$, there is one symmetric state that belongs to the symmetric subspace and the rest are part of the non-symmetric subspace, i.e the gaps to both symmetric and non-symmetric states is $\sim d$. The symmetric state in the Schwinger boson language corresponds to $(a^\dagger_{\uparrow})^n(a^{\dagger}_{\downarrow})^{d-n}\ket{0}$. Note that for these problems the ground state sector is spanned by $(a^{\dagger}_{\uparrow})^d\ket{0}$ and $(a^{\dagger}_{\downarrow})^d\ket{0}$.

Now let's the consider the $U(1)_\beta$ model which has a continuous $U(1)$ symmetry. Let's consider a very general model with such a continuous symmetry to illustrate how the difference between the two cases originated from the difference between a discrete and continuous symmetry associated with $\beta$. Consider a simple quantum XY model with $d$ phase variables $\theta_i$ and its conjugate variable $n_i$ with the Hamiltonian

\begin{equation}
    H = \frac{E_J}{2} \sum_{1 \leq i,j \leq d} (\theta_i - \theta_j)^2 + \frac{E_C}{2} \sum_{i=1}^{d} n_i^2
\end{equation}

The equations of motion for the above are $\dot{\theta}_i = E_c n_i$ and $\dot{n}_i = -E_j \sum_{j=1}^d (\theta_i - \theta_j)$. The corresponding eigenmodes for these are

\begin{equation}
    i \omega \tilde{\theta_i} = E_C \tilde{n}_i ;\; i \omega \tilde{n}_i = -E_J \sum_{j=1}^d (\tilde{\theta_i}-\tilde{\theta_j})
\end{equation}

For states in the non-symmetric sector $\sum_{j=1}^d \tilde{\theta_j} = 0$, which gives $\omega = \sqrt{dE_c E_j}$.Hence, the gap scales as $\sqrt{d}$, as claimed in the main text in \ref{sec_noiseanalysis}.

\bibliography{entmonbib.bib}
\end{document}